\documentclass{bmcart}

\usepackage{tikz}
\usetikzlibrary{shapes.geometric, arrows}
\usepackage{amssymb,amsfonts,amsmath}

\usepackage{graphicx}
\usepackage{epstopdf}

\usepackage[utf8]{inputenc}
\usepackage[T1]{fontenc}
\usepackage{color}
\usepackage{transparent}
\DeclareMathOperator{\tr}{Tr}

\usepackage[utf8]{inputenc} 
\usepackage[matrix,frame,arrow]{xypic}

\startlocaldefs
\endlocaldefs

\begin{document}

\begin{frontmatter}

\begin{fmbox}
\dochead{Research}


\title{Few-qubit quantum-classical simulation of strongly correlated lattice fermions}


\author[
   addressref={aff1},                   
   corref={aff1},                       
   email={j.kreula1@physics.ox.ac.uk}   
]{\inits{JM}\fnm{Juha M} \snm{Kreula}}
\author[
   addressref={aff2},
]{\inits{L}\fnm{Laura} \snm{García--Álvarez}}
\author[
   addressref={aff2},
]{\inits{L}\fnm{Lucas} \snm{Lamata}}
\author[
   addressref={aff3,aff4},
]{\inits{SR}\fnm{Stephen R} \snm{Clark}}
\author[
   addressref={aff2,aff5},
]{\inits{E}\fnm{Enrique} \snm{Solano}}
\author[
   addressref={aff1,aff6},
]{\inits{D}\fnm{Dieter} \snm{Jaksch}}


\address[id=aff1]{
  \orgname{Clarendon Laboratory, Department of Physics, University of Oxford}, 
  \street{Parks Road},
  \city{Oxford}                       %
  \postcode{OX1 3PU},                                
  \cny{UK}                                    
}
\address[id=aff2]{%
  \orgname{Department of Physical Chemistry, University of the Basque Country},
  \street{Apartado 644},
  \postcode{48080}
  \city{Bilbao},
  \cny{Spain}
}
\address[id=aff3]{%
  \orgname{Department of Physics, University of Bath},
  \street{Claverton Down},
  \city{Bath}
  \postcode{BA2 7AY}
  \cny{UK}
}
\address[id=aff4]{%
  \orgname{Max Planck Institute for the Structure and Dynamics of Matter},
  \city{Hamburg},
  \cny{Germany}
}
\address[id=aff5]{%
  \orgname{IKERBASQUE, Basque Foundation for Science},
  \street{Maria Diaz de Haro 3},
  \postcode{48013}
  \city{Bilbao},
  \cny{Spain}
}
\address[id=aff6]{%
  \orgname{Centre for Quantum Technologies, National University of Singapore},
  \street{3 Science Drive 2},
  \postcode{117543}
  \cny{Singapore}
}


\end{fmbox}


\begin{abstractbox}

\begin{abstract} 
We study a proof-of-principle example of the recently proposed hybrid quantum-classical simulation of strongly correlated fermion models in the thermodynamic limit. In a ``two-site'' dynamical mean-field theory (DMFT) approach we reduce the Hubbard model to an effective impurity model subject to self-consistency conditions. The resulting minimal two-site representation of the non-linear hybrid setup involves four qubits implementing the impurity problem, plus an ancilla qubit on which all measurements are performed. We outline a possible implementation with superconducting circuits feasible with near-future technology.
\end{abstract}


\begin{keyword}
\kwd{Quantum simulation}
\kwd{Dynamical mean-field theory}
\kwd{Superconducting circuits}
\end{keyword}


\end{abstractbox}
%

\end{frontmatter}



\section{Introduction}
Using highly controllable quantum devices to study other quantum systems, i.e., quantum simulation~\cite{feynman1982simulating,buluta2009quantum,cirac2012goals,johnson2014quantum}, offers a means to tackle strongly correlated fermion models that are intractable on classical computers. This is vital for understanding complex quantum materials~\cite{quantummat} with strong electronic correlations that exhibit a plethora of exciting physical phenomena of immediate technological interest. Examples of such effects include the Mott metal-insulator transition~\cite{mott1968metal,imada1998metal}, colossal magnetoresistance~\cite{ramirez1997colossal}, and high-temperature superconductivity~\cite{lee2006doping,schrieffer2007handbook}.

Classical numerical methods have limited ability to study even significantly simplified toy models of strongly correlated fermions. For instance, exact diagonalization faces exponential scaling with the system size, while quantum Monte Carlo methods~\cite{foulkes2001quantum,rubtsov2005continuous} are often crippled by the infamous fermionic sign problem~\cite{troyer2005computational}. Tensor network methods~\cite{vidal2003efficient,vidal2004efficient,verstraete2008matrix,cirac2009renormalization,schollwock2011density}  are powerful in one spatial dimension where they track strong correlations accurately. However, in higher dimensional systems, correlations tend to grow more quickly with system size, making these methods computationally challenging.

Another well-established approach to the study of strongly correlated fermionic lattice systems is dynamical mean-field theory (DMFT)~\cite{georges1996dynamical}. It reduces the complexity of the original problem, e.g., the Hubbard model~\cite{hubbard1963electron} \emph{in the thermodynamic limit}, by mapping it onto a simpler impurity problem that is subject to a self-consistency condition relating its properties to those of the original model. Since an impurity problem is local, the mapping corresponds to neglecting spatial fluctuations. In the limit of infinite spatial dimensions this mapping is exact, but for finite dimensions it is an approximation. Nonetheless for lattice geometries with a large coordination number, self-consistently solving the impurity problem can yield an accurate approximate solution to the original Hubbard problem.

The `impurity' itself consists of a single lattice site taken from the original problem, and so inherits on-site interactions from the Hubbard model. This impurity site is then immersed into a time-dependent, self-consistent mean-field with which it can dynamically exchange fermions. The mean-field thus attempts to model the rest of the lattice and by being dynamical can describe retardation phenomena. Overall the impurity problem can be represented by a Hamiltonian in which the interacting impurity site is coupled to a discrete set of non-interacting `bath' sites. The bath sites represent the mean-field and if there is an infinite number of them then the self-consistency condition is guaranteed to be satisfied. However, in practical implementations only a finite number of bath sites are used, which restricts the frequency resolution of the bath so self-consistency condition can only be fulfilled approximately. Nevertheless, many strongly-correlated features, e.g., the Mott transition, are still be captured correctly~\cite{georges1996dynamical}.

Although DMFT maps a Hubbard model to an impurity model this is still a non-trivial quantum many-body problem to solve because of the interactions at the impurity site. It is usually solved by classical numerical methods, e.g., specialised versions of those used to tackle the original problem, which attempt to keep track of the quantum correlations between impurity and bath sites. Again this limits the number of bath sites that can be treated accurately.

Here, we consider an alternative approach where the impurity problem is solved with a quantum simulator, thus avoiding many issues that are inherent to the classical methods. Quantum simulation of fermionic models has so far been mostly restricted to the analogue paradigm, especially with ultracold atoms in optical lattices~\cite{bloch2012quantum}. Digital simulation approaches, akin to universal quantum simulators \cite{lloyd1996universal}, have started to emerge in recent years, for example based on superconducting circuits~\cite{blais2004cavity,houck2012chip,barends2015digital,barends2016digitized}. The number of qubits in these digital simulators is, however, presently rather small. A direct implementation of the Hubbard model would suffer from severe finite size effects. It is nevertheless still possible for a digital quantum simulator with a restricted number of qubits to describe fermionic models directly in the thermodynamic limit when the DMFT approach is adopted.

To demonstrate this method we focus on the minimal incarnation of DMFT, the so-called ``two-site'' DMFT~\cite{potthoff2001two}, where the impurity model consists of one impurity site and only one bath site, both with local Hilbert space dimension four, subjected to two specially chosen self-consistency conditions. Since two-site DMFT considers only the smallest possible impurity model, the approach cannot match the accuracy of full DMFT, but it can still give a \emph{qualitatively} correct description of the infinite-dimensional Hubbard model, and its simplicity makes it a good starting point before advancing to more accurate schemes. For explicit details of two-site DMFT and its features compared to full DMFT we refer to Ref.~\cite{potthoff2001two}.

The two-site system corresponds to four qubits, two for the impurity site and two for the bath site, while a fifth, ancillary qubit is used for measurements. This number of qubits is readily available in current digital quantum simulator platforms, with IBM having made a five-qubit quantum processor available to the public~\cite{IBM}. A nine-qubit processor has already been demonstrated in superconducting circuits~\cite{kelly2015state,barends2015digital,barends2016digitized}. Trapped-ion technologies also allow for digital quantum simulations with up to six qubits~\cite{lanyon2011universal,innsbruck2016digital}. Being commensurate with current state-of-the-art technology is a further justification for studying this minimal model. Our scheme is readily generalisable to a larger number of qubits allowing for more accurate simulations and potentially offering an exponential speed-up over classical Hamiltonian-based DMFT methods~\cite{kreula2015coprocessor}.

The self-consistency conditions are taken care of iteratively in a classical feedback loop, which thus completes the non-linear, hybrid quantum-classical device we introduce. Dynamical mean-field simulations have already been proposed for such hybrid devices~\cite{bauer,kreula2015coprocessor}. Quantum gates similar to the ones needed in the two-site scheme have been used in demonstrating digital quantum simulation of fermionic models with superconducting circuits~\cite{las2015fermionic,barends2015digital}. We thus focus on superconducting circuits as a candidate platform, although, e.g., trapped ions~\cite{benhelm2008towards,lanyon2011universal,blatt2012quantum,casanova2012quantum} could also be considered.

This paper is organised as follows. In Section~\ref{sec:DMFT}, we further elucidate the framework of DMFT applied to the Hubbard model in infinite dimensions. Section~\ref{sec:2siteDMFT} introduces the two-site DMFT scheme in detail. Section~\ref{sec:algSIAM} discusses the implementation of this two-site scheme with special attention to superconducting circuits. In Section~\ref{sec:results}, we show the results of our analysis. We end with a summary in Section~\ref{sec:summ} and give an outline of the single-qubit interferometry measurement scheme in the  Appendix.

\section{Hubbard model in infinite dimensions and dynamical mean-field theory}\label{sec:DMFT}

A standard model to describe strongly correlated electron systems in thermodynamic equilibrium is the Hubbard Hamiltonian
\begin{align}\label{eq:hubbard}
\hat{H}=-t\sum_{\langle j, k \rangle \sigma} \left( \hat{c}^{\dagger}_{j,\sigma} \hat{c}_{k,\sigma} + \mathrm{H.c.} \right)+ U \sum_j \hat{n}_{j,\downarrow} \hat{n}_{j,\uparrow}.
\end{align}

In this model, electrons with spin projections $\sigma=\downarrow, \uparrow$ `hop' between adjacent lattice sites with tunnelling energy $t$. This process is described in the first term, where $\langle j, k \rangle$ denotes the sum over all nearest-neighbour sites $j$ and $k$, and $\hat{c}^{\dagger}_{j,\sigma}$ and $\hat{c}_{k,\sigma}$ denote the fermionic creation and annihilation operators, respectively. The electrons interact with on-site Coulomb repulsion $U>0$, described in the latter term by the product of the local number operators $\hat{n}_{j,\downarrow}=\hat{c}^{\dagger}_{j,\downarrow}\hat{c}_{j,\downarrow}$  and $\hat{n}_{j,\uparrow}=\hat{c}^{\dagger}_{j,\uparrow}\hat{c}_{j,\uparrow}$.

Here, we consider the paramagnetic Hubbard model in an infinite-dimensional Bethe lattice in the thermodynamic limit at zero temperature. This setup has very simple self-consistency relations, which makes it an ideal test-bed for a proof-of-principle demonstration of a hybrid quantum-classical scheme.

\begin{figure}[ht!]
\centerline{
\includegraphics[scale=0.12]{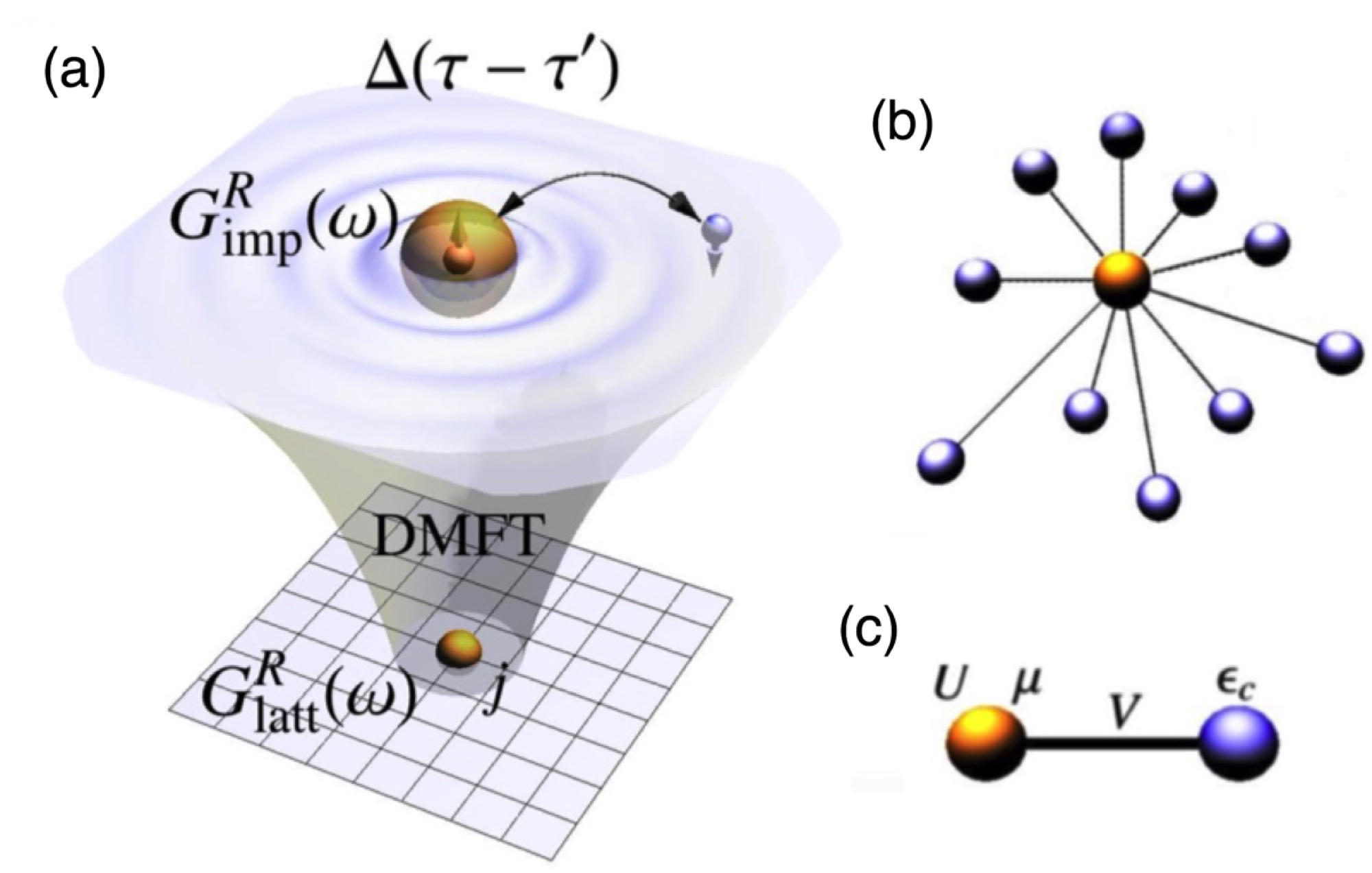}
}
\caption{\csentence{Dynamical mean-field theory}. (a) DMFT neglects spatial fluctuations around a single lattice site $j$ and replaces the rest of the lattice with an effective mean-field $\Delta(\tau-\tau')$ with which the isolated site dynamically exchanges fermions, subject to the self-consistency condition $G^R_{\rm imp}(\omega)=G^R_{{\rm latt},jj}(\omega)$. Here, $G^R_{\rm imp}(\omega)$ is the impurity Green function and $G^R_{{\rm latt},jj}(\omega)$ is the local part of the lattice Green function. (b) In Hamiltonian-based DMFT methods, one considers an impurity model which describes the local part of the Hubbard model directly and represents the mean-field as a set of non-interacting bath sites that are connected to the central, interacting impurity site. (c) The minimal representation of DMFT involves the impurity site, with on-site interaction $U$ and chemical potential $\mu$, coupled via the hybridization energy $V$ to only one bath site. The bath has on-site energy $\epsilon_c$ that corresponds to the mean-field $\Delta(\tau-\tau')$ and is subject to two self-consistency conditions.}
\label{fig:dmft}
\end{figure}

The DMFT approach~\cite{georges1996dynamical} to solving this model consists in neglecting spatial fluctuations around a single lattice site and replacing the rest of the many-body lattice in the thermodynamic limit by a time-translation-invariant, self-consistent mean-field $\Delta(\tau-\tau')$ (or $\Delta(\omega)$ in the frequency domain), as illustrated in Fig.~\ref{fig:dmft}a. The isolated lattice site can dynamically exchange fermions with the mean-field at time instants $\tau'$ and $\tau$. This allows one to include retardation effects that are important in the presence of strong correlations.  In short, the dynamical mean-field approach reduces the complexity of the full Hubbard model to an effective single-site system which is a slightly more benign many-body problem to solve. In infinite dimensions, DMFT becomes exact as the irreducible self-energy of the lattice model becomes strictly local in space, $\Sigma_{{\rm latt},jk}(\omega)=\delta_{jk}\Sigma_{{\rm latt},jj}(\omega)$, and its skeleton diagrams agree with those of a single-site, or impurity, model~\cite{georges1996dynamical}.

The solution of the effective single-site, or impurity, problem also yields the solution of the infinite-dimensional Hubbard model due to the self-consistency condition. This leads to the retarded single-particle impurity Green function in the frequency domain being given by
\begin{align}\label{eq:Gfreq}
G^R_{\rm imp}(\omega)=\frac{1}{\omega+\mu-\Delta(\omega)-\Sigma_{\rm imp}(\omega)},
\end{align}
where $\mu$ is the chemical potential, and $\Sigma_{\rm imp}(\omega)$ denotes the impurity self-energy. We set $\hbar=1$ throughout the paper. The impurity Green function describes the response of the many-body system after a localized removal or addition of a particle on the impurity site and is defined in the time domain and at zero temperature as
\begin{align}\label{eq:Gtime}
iG^R_{\rm imp}(\tau)=\theta(\tau) \langle \{ \hat{c}_{\sigma}(\tau), \hat{c}^{\dagger}_{\sigma}(0)  \}  \rangle,
\end{align}
where $i$ is the imaginary unit, $\tau$ is real time, $\{\cdot,\cdot\}$ denotes the anticommutator, $\theta(\tau)$ is the Heaviside step function, and the average is computed in the ground-state $|GS\rangle$ of the impurity model. The fermionic creation and annihilation operators are given in the Heisenberg picture. In the paramagnetic phase the Green function is spin symmetric and we therefore only need to work out $G^R_{\rm imp}(\omega)$ for one spin configuration.

The initially unknown mean-field $\Delta(\omega)$ has to be chosen such that $G^R_{\rm imp}(\omega)$ matches the local part of the retarded lattice Green function $G^R_{{\rm latt},jj}(\omega)$, i.e.,
\begin{align}\label{eq.selfcon}
G^R_{\rm imp}(\omega)=G^R_{{\rm latt},jj}(\omega),
\end{align}
where $j$ is the (randomly chosen) lattice site from which the removal or addition of a particle occurs in the translationally invariant lattice model. The DMFT self-consistency condition Eq.~\eqref{eq.selfcon} implies
\begin{align}\label{eq:sigmas}
\Sigma_{\rm imp}(\omega)=\Sigma_{{\rm latt},jj}(\omega),
\end{align}
i.e., the impurity self-energy matches the local self-energy of the Hubbard model in the infinite-dimensional Bethe lattice.

In the general case, the DMFT self-consistency loop is iterated as follows (see also Ref.~\cite{georges1996dynamical}). (i) First, guess the local self-energy $\Sigma_{{\rm latt},jj}(\omega)$. (ii) The local lattice Green function can be computed as $G^R_{{\rm latt},jj}(\omega)=\int_{-\infty}^{\infty} d\epsilon \, \rho_0(\epsilon)/\left[\omega +\mu-\epsilon-\Sigma_{{\rm latt},jj}(\omega)\right]$, where  $\rho_0(\epsilon)=\sqrt{4{t^*}^2-\epsilon^2}/2\pi{t^*}^2$ is the non-interacting density of states of a Bethe lattice. The constant ${t^*}$ emerges from the requirement that the Hubbard hopping needs to be scaled as $t \sim t^*/\sqrt{z}$ to avoid a diverging kinetic energy per lattice site in the limit of infinite coordination, $z \rightarrow \infty$ \cite{georges1996dynamical}. (iii) With Eqs.~\eqref{eq.selfcon} and~\eqref{eq:sigmas}, we obtain $\Delta(\omega)$ from Eq.~\eqref{eq:Gfreq} and the impurity model is then defined. (iv) Compute the impurity Green function and obtain the impurity self-energy $\Sigma_{\rm imp}(\omega)$. There are several means to do this~\cite{georges1996dynamical}. (v) Set $\Sigma^{\rm new}_{{\rm latt},jj}(\omega)=\Sigma_{\rm imp}(\omega)$. (vi) Check if the self-energy has converged. If not, go to step (ii) and repeat.

Once self-consistent, the solution of the impurity problem then gives access to local single-particle properties of the original lattice model. For example, the local lattice spectral function is given by
\begin{align*}
{A_{{\rm latt},jj}(\omega)=-{\rm Im}[G^R_{{\rm latt},jj}(\omega+i\eta)]/\pi=-{\rm Im}[G^R_{{\rm imp}}(\omega+i\eta)]/\pi},
\end{align*}
where $\eta$ is a positive infinitesimal.

In Hamiltonian-based impurity solvers, one parameterizes $\Delta(\omega)$ by a set of bath sites (see Fig.~\ref{fig:dmft}b). For any finite number of bath sites, the self-consistency condition~\eqref{eq.selfcon} can only be approximately satisfied and in the extreme ``two-site'' DMFT it turns out to be more suitable to reformulate Eq.~\eqref{eq.selfcon} in a manner specially focused on this minimal representation ~\cite{potthoff2001two} (see Section~\ref{sec:2siteDMFT}). Note that two-site DMFT is only able to provide a qualitatively correct description of the Hubbard model even in infinite dimensions~\cite{potthoff2001two}.

\section{Quantum simulator based on two-site DMFT}\label{sec:2siteDMFT}
In terms of the single-impurity Anderson model (SIAM), the smallest impurity problem involves one fermionic site corresponding to the impurity and only one fermionic site corresponding to the entire mean-field as described in the previous section. Since two qubits are needed to encode the local Hilbert space of a fermionic site, we only require four physical qubits to implement this representation in the lab. The SIAM Hamiltonian for only one bath site reads
\begin{align}\label{eq:HSIAM}
\hat{H}_{\mathrm{SIAM}}=&U\hat{n}_{1\downarrow}\hat{n}_{1\uparrow}-\mu\sum_\sigma \hat{n}_{1\sigma}+ \sum_{\sigma} \epsilon_c \hat{c}^{\dagger}_{2\sigma}\hat{c}_{2\sigma}+\sum_{\sigma} V\left(\hat{c}^{\dagger}_{1\sigma}\hat{c}_{2\sigma} + \mathrm{H.c.} \right).
\end{align}

Here, $U$ is the Hubbard interaction at the impurity site 1, and $\mu$ is the chemical potential that controls the electron filling in the grand canonical ensemble. Furthermore, $\epsilon_c$ and $V$ describe the on-site energy of the non-interacting bath site 2 and hybridization between the impurity and the bath site, respectively, and give the mean-field as
\begin{align}
\Delta(\omega)=\frac{V^2}{\omega-\epsilon_c}.
\end{align}

See Fig.~\ref{fig:dmft}c for illustration of the two-site SIAM. The parameters $\epsilon_c$ and $V$ are initially unknown and they need to be determined iteratively such that two self-consistency conditions are satisfied. For details of the derivation and motivation of these conditions we refer to Ref.~\cite{potthoff2001two}.

The first condition is that the electron filling $n_{\mathrm{imp}}$ of the impurity site and the filling $n=\langle n_{j\downarrow} \rangle + \langle n_{j\uparrow} \rangle$ of the lattice model match, i.e.,
\begin{align}
n_{\mathrm{imp}} \equiv n .
\end{align}

The second self-consistency condition is given by
\begin{align}\label{eq:VZ}
V^2=\mathcal{Z}M_2^{(0)}= \mathcal{Z}\int_{-\infty}^{\infty} d\epsilon \, \epsilon^2 \rho_0(\epsilon)=\mathcal{Z}{t^*}^2,
\end{align}
where quasiparticle weight reads
\begin{align}\label{eq:Z}
\mathcal{Z}=\left[1- \frac{d{\rm Re}[\Sigma_{\rm imp}(\omega+i\eta)]}{d\omega}\Big|_{\omega=0} \right]^{-1}.
\end{align}
In Eq.~\eqref{eq:VZ}, $M_2^{(0)}$ is the second moment of the non-interacting density of states, and the final equality follows from the semicircular density of states of the Bethe lattice.
\begin{figure}[ht!]
\centerline{
\includegraphics[scale=1]{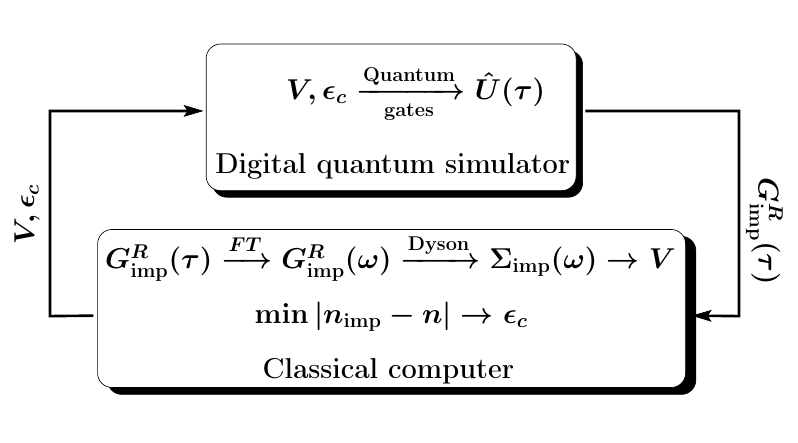}
}
\caption{\csentence{Non-linear hybrid quantum-classical scheme.}
     A digital quantum simulator works in conjunction with a classical feedback loop to perform a proof-of-principle demonstration of a two-site DMFT calculation.}
\label{fig:superloop}
\end{figure}
\subsection{Two-site DMFT protocol}
The hybrid quantum-classical device implementing two-site DMFT consists of a few-qubit digital quantum simulator in which the impurity Green function is measured and of a classical feedback loop in which the parameters of the two-site SIAM are updated. The two-site DMFT protocol is summarized in Fig.~\ref{fig:superloop} and proceeds as follows (see also Ref.~\cite{potthoff2001two}).

\begin{itemize}
\item[1.] First fix $U$ and $\mu$ to the desired values in the SIAM and set the unknown parameters $\epsilon_c$ and $V$ equal to an initial guess.
\item[2.]  Measure the interacting Green function $iG^R_{\mathrm{imp}}(\tau)$. This can be done using, e.g., single-qubit interferometry (see details in the Appendix).
\item[3.] After Fourier-transforming the impurity Green function, the impurity self-energy is obtained classically from the Dyson equation
\begin{align}\label{eq:dyson}
\Sigma_{\rm imp}(\omega) = {G}_{\rm imp}^{R(0)}(\omega)^{-1}-G^R_{\mathrm{imp}}(\omega)^{-1}.
\end{align}

Here, the non-interacting impurity Green function is given by
\begin{align}\label{eq:nonintGF}
{G}_{\rm imp}^{R(0)}(\omega)^{-1}=\omega+\mu-\Delta(\omega).
\end{align}

From the derivative of the self-energy one obtains the quasiparticle weight $\mathcal{Z}$ which directly yields the updated hopping parameter $V$ via Eq.~\eqref{eq:VZ}. The update for $\epsilon_c$ is found by minimizing the difference $|n_{\rm imp}-n|$~\cite{potthoff2001two}.
\item[4.] Steps 2 and 3 need to be repeated until $V$ and $\epsilon_c$ are self-consistent, and $n_{\rm imp}=n$.
\end{itemize}
The self-consistent Green function $G^R_{\rm imp}(\omega)$ and self-energy $\Sigma_{\rm imp}(\omega)$ thus obtained are used to calculate approximations to local single-particle properties of the Hubbard model.

\section{Quantum algorithm for the single-impurity Anderson model with superconducting circuits}\label{sec:algSIAM}
Here, we consider the quantum gates of the digital quantum simulator part in Fig.~\ref{fig:superloop}, with special focus on superconducting circuits as the platform of choice~\cite{las2015fermionic,barends2015digital,barends2016digitized}.

\subsection{Jordan--Wigner transformation of the SIAM}
To implement the two-site SIAM with qubits, the fermionic creation and annihilation operators need to be mapped onto tensor products of spin operators which then act on the qubits via quantum gates. In order to obtain as simple quantum gates as possible in Section~\ref{sec:quantumevol} and in the Appendix, we consider an ordering of the qubits where the first two qubits encode the spin $\downarrow$ for both fermionic sites while the last two correspond to spin $\uparrow$. This is achieved via the Jordan--Wigner transformation given explicitly as
\begin{eqnarray}
\hat{c}^{\dagger}_{1\downarrow}&=&\hat{\sigma}_{1}^-\;=\;\; \frac{1}{2}\left(\hat{\sigma}_1^x-i\hat{\sigma}_1^y \right),\\
\hat{c}^{\dagger}_{2\downarrow}&=&\hat{\sigma}^z_1 \hat{\sigma}_{2}^-\;=\;\;\frac{1}{2}\hat{\sigma}^z_1 \left(\hat{\sigma}_2^x-i\hat{\sigma}_2^y \right),\\
\hat{c}^{\dagger}_{1\uparrow}&=&\hat{\sigma}^z_1\hat{\sigma}^z_2 \hat{\sigma}_{3}^-\;=\;\;\frac{1}{2}\hat{\sigma}^z_1 \hat{\sigma}^z_2 \left(\hat{\sigma}_3^x-i\hat{\sigma}_3^y \right),\\
\hat{c}^{\dagger}_{2\uparrow}&=&\hat{\sigma}^z_1  \hat{\sigma}^z_2  \hat{\sigma}^z_3 \hat{\sigma}_{4}^-\;=\;\;\frac{1}{2}\hat{\sigma}^z_1 \hat{\sigma}^z_2  \hat{\sigma}^z_3 \left(\hat{\sigma}_4^x-i\hat{\sigma}_4^y \right),
\end{eqnarray}
and $\hat{c}_{j\sigma}=\left(\hat{c}_{j\sigma}^{\dagger} \right)^{\dagger}$. With this mapping the hybridization terms  in the SIAM described in Eq.~\eqref{eq:HSIAM} transform into
\begin{align}
V\left(\hat{c}^{\dagger}_{1\downarrow}\hat{c}_{2\downarrow} + \mathrm{H.c.} \right)=\frac{V}{2}\left(\hat{\sigma}_1^x \hat{\sigma}_2^x + \hat{\sigma}_1^y  \hat{\sigma}_2^y \right),
\end{align}
and
\begin{align}
V\left(\hat{c}^{\dagger}_{1\uparrow}\hat{c}_{2\uparrow} + \mathrm{H.c.} \right)=\frac{V}{2}\left(\hat{\sigma}_3^x  \hat{\sigma}_4^x + \hat{\sigma}_3^y  \hat{\sigma}_4^y \right).
\end{align}

The number operators become
\begin{eqnarray}
\hat{n}_{1\downarrow} &=& \frac{1}{2}\left(\hat{I} - \hat{\sigma}_1^z \right),\\
\hat{n}_{2\downarrow} &=& \frac{1}{2}\left(\hat{I} - \hat{\sigma}_2^z \right),\\
\hat{n}_{1\uparrow} &=& \frac{1}{2}\left(\hat{I} - \hat{\sigma}_3^z \right),\\
\hat{n}_{2\uparrow} &=& \frac{1}{2}\left(\hat{I} - \hat{\sigma}_4^z \right),
\end{eqnarray}
and thus the interaction term can be written as
\begin{align}
U\hat{n}_{1\downarrow}\hat{n}_{1\uparrow}=\frac{U}{4}(\hat{\sigma}_1^z  \hat{\sigma}_3^z - \hat{\sigma}_1^z  - \hat{\sigma}^z_3),
\end{align}
up to a constant. The total Hamiltonian then reads
\begin{align}\label{eq:SIAMspin}
\hat{H}_{\rm SIAM}=& \frac{U}{4} \left( \hat{\sigma}^z_1 \hat{\sigma}^z_3  - \hat{\sigma}^z_1 - \hat{\sigma}^z_3 \right) + \frac{\mu}{2} \left( \hat{\sigma}^z_1 + \hat{\sigma}^z_3 \right) - \frac{\epsilon_{c}}{2} \left( \hat{\sigma}^z_2 + \hat{\sigma}^z_4 \right) \nonumber \\
&+ \frac{V}{2} \left( \hat{\sigma}^x_1\hat{\sigma}^x_2 + \hat{\sigma}^y_1\hat{\sigma}^y_2 + \hat{\sigma}^x_3\hat{\sigma}^x_4 + \hat{\sigma}^y_3\hat{\sigma}^y_4 \right),
\end{align}
where we have dropped constant terms.

\subsection{Quantum gates in superconducting circuits}\label{sec:gatessuper}
We now consider how the Jordan--Wigner transformed SIAM in Eq.~\eqref{eq:SIAMspin} can be implemented in an experimental arrangement based on superconducting circuits. We present two alternative approaches. The first one couples the qubits with a transmission line resonator, which leads to the so-called $XY$ gate between the qubits. The second approach is the Controlled-Z$_\phi$ (CZ$_\phi$) gate, which can be obtained via a capacitive coupling of nearest-neighbour transmon qubits without using a resonator. These CZ$_\phi$ gates have been implemented with high fidelities of above 99\% for a variant of transmon qubits called `X-mon' qubits \cite{barends2014superconducting}.

\paragraph{$XY$ gates with resonators ---} The basic Hamiltonian coupling a set of qubits to the resonator has the form of a detuned Jaynes-Cummings model. By adiabatically eliminating the resonator one obtains, when the resonator is in the vacuum state, the well-known $XY$ model for a pair of qubits $l$ and $m$ as
\begin{equation}
\hat{H}_{XY}=\frac{g_l g_m}{2\Delta}(\hat{\sigma}^x_l\hat{\sigma}^x_m+\hat{\sigma}^y_l\hat{\sigma}^y_m).
\end{equation}

Here, $\Delta$ is the detuning between the qubit level-spacing and the resonator mode, $g_l$ is the coupling constant between qubit $l$ and the resonator, and $\hat{\sigma}^x$ and $\hat{\sigma}^y$ are Pauli operators. The $XY$ gate is universal for quantum computation and simulation in combination with single qubit gates, and is the natural interaction customarily employed in superconducting circuits.

\paragraph{CZ-$\phi$ gates with capacitive couplings ---} To perform the CZ-$\phi$ gate, one qubit is kept at a fixed frequency while the other carries out an adiabatic trajectory near an appropriate resonance of the two-qubit states. By varying the amplitude of this trajectory one can tune the conditional phase $\phi$. The unitary for the CZ$_\phi$ is given by
\begin{align}\label{eq:cz}
{\rm CZ}_{\phi}=\left( \begin{array}{cccc}
1 & 0 & 0 & 0 \\ 0 & 1 & 0 & 0 \\ 0 & 0 & 1 & 0 \\ 0 & 0 & 0 & e^{i\phi}
\end{array} \right).
\end{align}

\subsection{Quantum gate decomposition of the time-evolution operator}\label{sec:quantumevol}
In order to use quantum gates for time-evolution, we utilize a Trotter decomposition of the time-evolution operator corresponding to $\hat{H}_{\rm SIAM}$ in Eq.~\eqref{eq:SIAMspin}. The first order Trotter expansion is given by
\begin{align}\label{eq:TrotterU}
\hat{U}(\tau)=e^{-i\hat{H}_{\rm SIAM} \tau} &\approx   \bigg(e^{-i \frac{V}{2} (\hat{\sigma}^{x}_1 \hat{\sigma}^{x}_2 + \hat{\sigma}^{y}_1 \hat{\sigma}^{y}_2) \frac{\tau}{N}} e^{-i \frac{V}{2} (\hat{\sigma}^{x}_3 \hat{\sigma}^{x}_4 + \hat{\sigma}^{y}_3 \hat{\sigma}^{y}_4) \frac{\tau}{N}} e^{-i \frac{U}{4} \hat{\sigma}^{z}_1 \hat{\sigma}^{z}_3 \frac{\tau}{N}} \nonumber \\
&  \times \ e^{i \big(\frac{U}{4}-\frac{\mu}{2}\big) \hat{\sigma}^{z}_1 \frac{\tau}{N}} e^{i \big(\frac{U}{4}-\frac{\mu}{2}\big) \hat{\sigma}^{z}_3 \frac{\tau}{N}} e^{i \frac{\epsilon_c}{2} \hat{\sigma}^{z}_2 \frac{\tau}{N}} e^{i \frac{\epsilon_c}{2} \hat{\sigma}^{z}_4 \frac{\tau}{N}} \bigg)^{N}.
\end{align}

Here, $N$ is the number of Trotter (i.e., time) steps and $\frac{\tau}{N}$ is the size of the time step. In what follows, we use the two alternative approaches for quantum gates outlined in Section~\ref{sec:gatessuper} to implement Eq.~\eqref{eq:TrotterU}.

\begin{figure}
\centerline{
\includegraphics[scale=0.25]{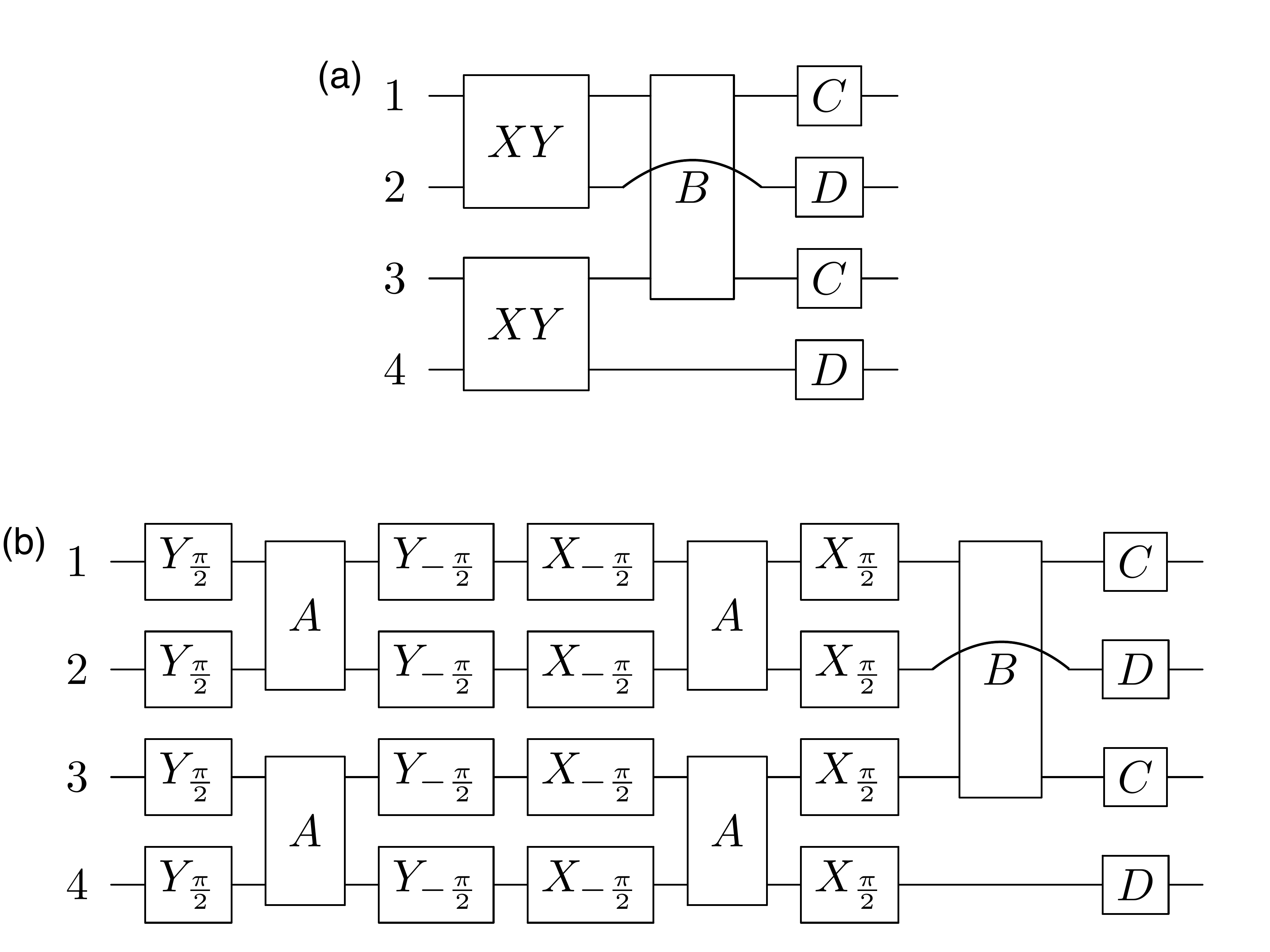}
}
     \caption{\csentence{Quantum gates for one Trotter step.} A single Trotter step is shown for (a) the $XY$ method and (b) the CZ method. Here, $B$ is the entangling gate $B = \exp\big(-i\frac{U}{4}\hat{\sigma}^{z}_1 \hat{\sigma}^{z}_3 \frac{\tau}{N}\big)$, $A$ is a two-qubit gate given by ${A = \exp\big(-i\frac{V}{2}\hat{\sigma}^{z}_l \hat{\sigma}^{z}_m \frac{\tau}{N}\big)}$, acting on qubits $l$ and $m$, and the quantum gates $C$ and $D$ are single qubit $\hat{\sigma}^{z}$-gates, given by $C = \exp\left[i\big(\frac{U}{4}-\frac{\mu}{2}\big)\hat{\sigma}^{z}_l\frac{\tau}{N}\right]$ and $D = \exp\big(i\frac{\epsilon_c}{2}\hat{\sigma}^{z}_l\frac{\tau}{N}\big)$, acting on qubit $l$. Finally, $X_{\phi}$ and $Y_{\phi}$ are $\phi$-rotations along the $x$ and $y$ axis, respectively.}
     \label{fig:xycz}
\end{figure}

\paragraph{$XY$ gates ---}
As shown in Section~\ref{sec:gatessuper}, the $XY$ gate, given by the expression $XY = {\exp\left[-i\frac{V}{2}(\hat{\sigma}^{x}_l \hat{\sigma}^{x}_m + \hat{\sigma}^{y}_l \hat{\sigma}^{y}_m) \frac{\tau}{n}\right]}$, naturally appears when considering the use of a resonator quantum bus~\cite{lasheras2014digital}. The quantum circuit for a single Trotter step with these gates is shown in Fig.~\ref{fig:xycz}a.

\paragraph{CZ-$\phi$ gates ---}
To be able to utilize the CZ-$\phi$ gates, we write the time-evolution operator in Eq.~\eqref{eq:TrotterU} in terms of $\hat{\sigma}^{z}$-$\hat{\sigma}^{z}$ (ZZ) interactions, taking into account that
\begin{align}
\hat{\sigma}^{x}_l \hat{\sigma}^{x}_m = \mathcal{R}^{(l)}_{y} (\tfrac{\pi}{2}) \hat{\sigma}^{z}_l \mathcal{R}^{(l)}_{y} (-\tfrac{\pi}{2}) \mathcal{R}^{(m)}_{y} (\tfrac{\pi}{2}) \hat{\sigma}^{z}_m \mathcal{R}^{(m)}_{y} (-\tfrac{\pi}{2}),
\end{align}
and
\begin{align}
\hat{\sigma}^{y}_l \hat{\sigma}^{y}_m = \mathcal{R}^{(l)}_{x} (-\tfrac{\pi}{2}) \hat{\sigma}^{z}_l \mathcal{R}^{(l)}_{x} (\tfrac{\pi}{2}) \mathcal{R}^{(m)}_{x} (-\tfrac{\pi}{2}) \hat{\sigma}^{z}_m \mathcal{R}^{(m)}_{x} (\tfrac{\pi}{2}),
\end{align}
where $\mathcal{R}^{(l)}_{\alpha} (\theta) = \exp(-i\frac{\theta}{2} \hat{\sigma}_l^{\alpha})$ is the rotation along the $\alpha$-axis of the $l$th qubit. Note that in the computational basis, one can write, e.g.,
\begin{align}
\exp \left(-i\,  \frac{\phi}{2} \hat{\sigma}^z_1  \hat{\sigma}^z_2 \right)=\left( \begin{array}{cccc}
1 & 0 & 0 & 0 \\ 0 & e^{i\phi} & 0 & 0 \\ 0 & 0 & e^{i\phi} & 0 \\ 0 & 0 & 0 & 1
\end{array} \right),
\end{align}
where we have neglected global phases. Thus, we have the decomposition
\begin{align}\label{eq:czaux}
&\exp \left(-i\, \frac{\phi}{2} \hat{\sigma}^z_1  \hat{\sigma}^z_2 \right)=\mathcal{R}^{(1)}_x(\pi)\, {\rm CZ}_{\phi}\, \mathcal{R}^{(1)}_x(\pi) \mathcal{R}^{(2)}_x(\pi)\, {\rm CZ}_{\phi}\,  \mathcal{R}^{(2)}_x(\pi),
\end{align}
where the tunable ${\rm CZ}_{\phi}$-gate is given by Eq.~\eqref{eq:cz}.

The time-evolution operator in Eq.~\eqref{eq:TrotterU} in terms of ZZ interactions is given by
\begin{align}
&\hat{U}(\tau)=e^{-i\hat{H}_{\rm SIAM} \tau} \approx  \bigg(\mathcal{R}^{(1234)}_y(\tfrac{\pi}{2}) e^{-i \frac{V}{2} \hat{\sigma}^{z}_1 \hat{\sigma}^{z}_2 \frac{\tau}{N}} e^{-i \frac{V}{2} \hat{\sigma}^{z}_3 \hat{\sigma}^{z}_4 \frac{\tau}{N}} \mathcal{R}^{(1234)}_y(-\tfrac{\pi}{2}) \nonumber \\
& \times \mathcal{R}^{(1234)}_x(-\tfrac{\pi}{2}) e^{-i \frac{V}{2} \hat{\sigma}^{z}_1 \hat{\sigma}^{z}_2 \frac{\tau}{N}} e^{-i \frac{V}{2} \hat{\sigma}^{z}_3 \hat{\sigma}^{z}_4 \frac{\tau}{N}} \mathcal{R}^{(1234)}_x(\tfrac{\pi}{2}) \nonumber \\
& \times \ e^{-i \frac{U}{4} \hat{\sigma}^{z}_1 \hat{\sigma}^{z}_3 \frac{\tau}{N}} e^{i \big(\frac{U}{4}-\frac{\mu}{2}\big) \hat{\sigma}^{z}_1 \frac{\tau}{N}} e^{i \big(\frac{U}{4}-\frac{\mu}{2}\big) \hat{\sigma}^{z}_3 \frac{\tau}{N}} e^{i \frac{\epsilon_c}{2} \hat{\sigma}^{z}_2 \frac{\tau}{N}} e^{i \frac{\epsilon_c}{2} \hat{\sigma}^{z}_4 \frac{\tau}{N}} \bigg)^{N},
\end{align}
\begin{figure}
\centerline{
\includegraphics[scale=0.25]{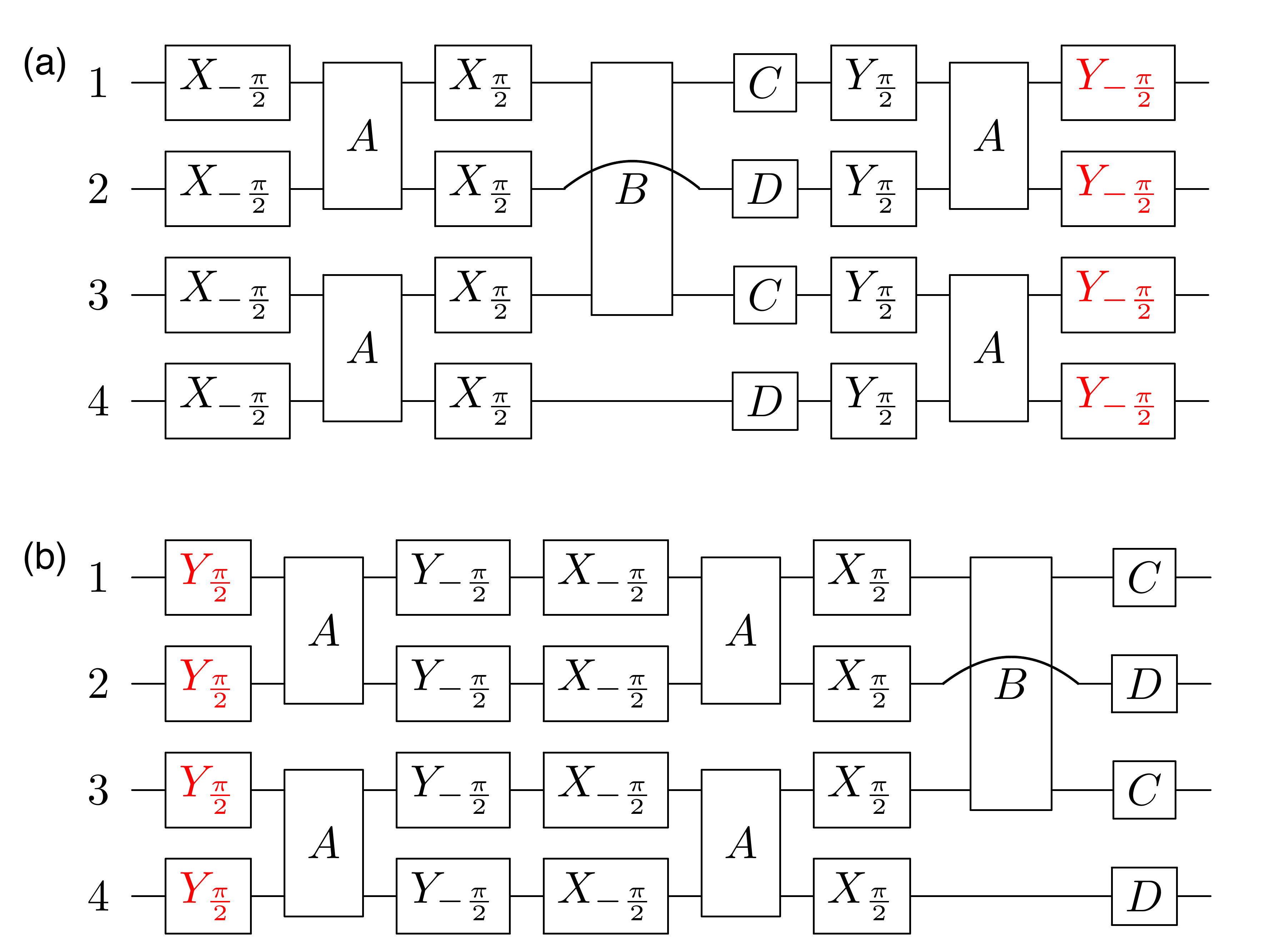}
}
     \caption{\csentence{Reordering of quantum gates.} The ordering of gates shown for (a) an odd Trotter step and (b) an even Trotter step in the CZ method. The gates depicted in red can be omitted as they cancel out during a sequence of time steps.}
     \label{fig:oddeven}
\end{figure}
\noindent
where $\mathcal{R}^{(1234)}_\alpha(\phi)=\mathcal{R}^{(1)}_{\alpha} (\phi)\mathcal{R}^{(2)}_{\alpha} (\phi) \mathcal{R}^{(3)}_{\alpha} (\phi)\mathcal{R}^{(4)}_{\alpha}(\phi) $. The sequence of gates for one Trotter step is depicted in Fig.~\ref{fig:xycz}b.

A single Trotter step contains 5~${\rm ZZ}$ two-qubit gates (corresponding to the $A$ and $B$ gates in Fig.~\ref{fig:xycz}b) between nearest-neighbour qubits, 2~${\rm SWAP}$ gates (for the $B$ gate which acts on qubits 1 and 3), and 20~single-qubit rotations. We note that a SWAP-gate amounts to three CZ-$\phi$ gates, and a ZZ-gate amounts to two CZ-$\phi$ gates (see Eq.~\eqref{eq:czaux}). This number can be optimised further if we consider different orderings for odd and even Trotter steps as in Fig.~\ref{fig:oddeven}, such that subsequent gates may be suppressed. This reorganisation of interactions does not in principle affect the Trotter error. Hence, for a pair of Trotter steps, the number of gates is reduced, and we may only consider 10~${\rm ZZ}$ two-qubit gates between nearest-neighbour qubits, 4~${\rm SWAP}$ gates, and 32~single-qubit rotations.

\section{Results}\label{sec:results}

We focus on the half-filled case, i.e., $\mu=\frac{U}{2}$ and $\epsilon_c=0$, which requires the least amount of quantum gates, since the $C$ and $D$ gates in Section~\ref{sec:algSIAM} vanish. Note that since the value of $\epsilon_c$ is fixed in this case, it need not be updated in the self-consistency loop. We use $t^*$, the Hubbard hopping in infinite dimensions, as our unit of energy, hence time $\tau$ is measured in units of $1/t^*$. Note that $\tau$ refers here to the time in the evolution operator $\hat{U}(\tau)$, not to the actual time to run the experiment.

\begin{figure}
\centerline{\includegraphics[scale=1.45]{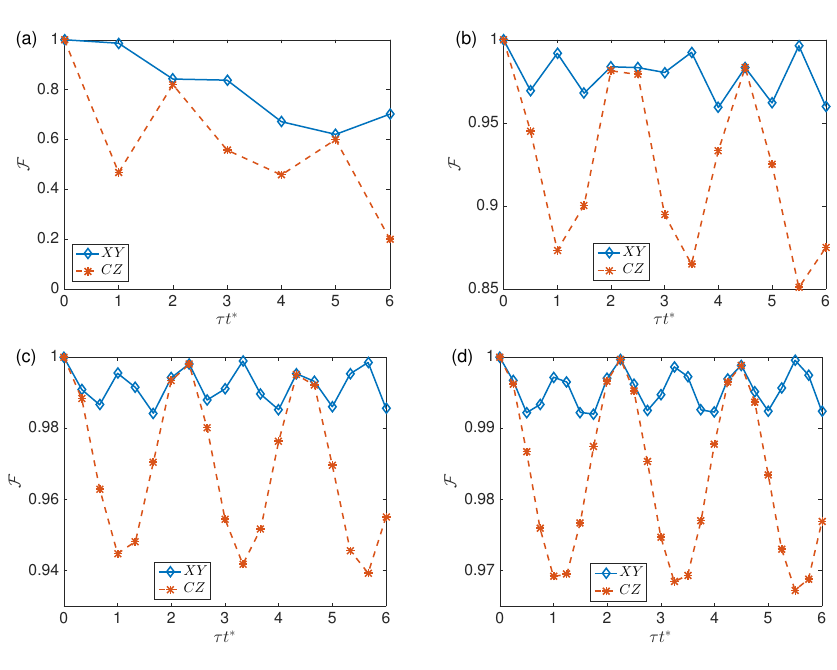}}
\caption{\csentence{Time-evolution of the state fidelity.} State fidelities $\mathcal{F}=| \langle \Psi(\tau) | \Psi_T(\tau) \rangle |^2$ using the $XY$ method (blue diamonds, line is to guide the eye) and $CZ$ gates (red stars, dashed line is to guide the eye) obtained with (a) 6, (b) 12, (c) 18, and (d) 24 Trotter steps up to time $\tau=6/t^*$. We set $U=4t^*$ and $V=t^*$.}
\label{fig:fid}
\end{figure}

We show in Fig.~\ref{fig:fid} the state fidelities $\mathcal{F}=| \langle \Psi(\tau) | \Psi_T(\tau) \rangle |^2$, where $|\Psi(\tau)\rangle$ denotes the state obtained with exact time-evolution using the full, non-Trotterized operator $\hat{U}(\tau)=\exp(-i\tau\hat{H}_{\rm SIAM})$ corresponding to the two-site SIAM in Eq.~\eqref{eq:HSIAM}, and $| \Psi_T(\tau) \rangle$ is the state evolved using either the $XY$ or CZ-$\phi$ quantum gates, for various Trotter steps $N$ up to time $\tau=6/t^*$. Note that the number of qubits corresponding to the two-site SIAM is fixed, leaving only $N$ as the parameter to be varied for increased accuracy. We use the initial state $|\Psi(\tau=0)\rangle=\hat{c}^{\dagger}_{1\downarrow}|GS\rangle/||\hat{c}^{\dagger}_{1\downarrow}|GS\rangle||$, where $|GS\rangle$ is the ground-state of the two-site SIAM in Eq.~\eqref{eq:HSIAM}, which is a relevant state for obtaining the impurity Green function at zero temperature (see Eq.~\eqref{eq:Gtime}).  As expected, using $XY$ gates displays superior fidelities, since CZ-$\phi$ gates require an extra factorization of the hybridization term (see Section~\ref{sec:algSIAM}). For $N=24$ steps, the state fidelity using $XY$ gates remains over 99\% throughout the evolution. In what follows, we use only $XY$ gates for the time-evolution for concreteness.

\begin{figure}
\centerline{\includegraphics[scale=1.45]{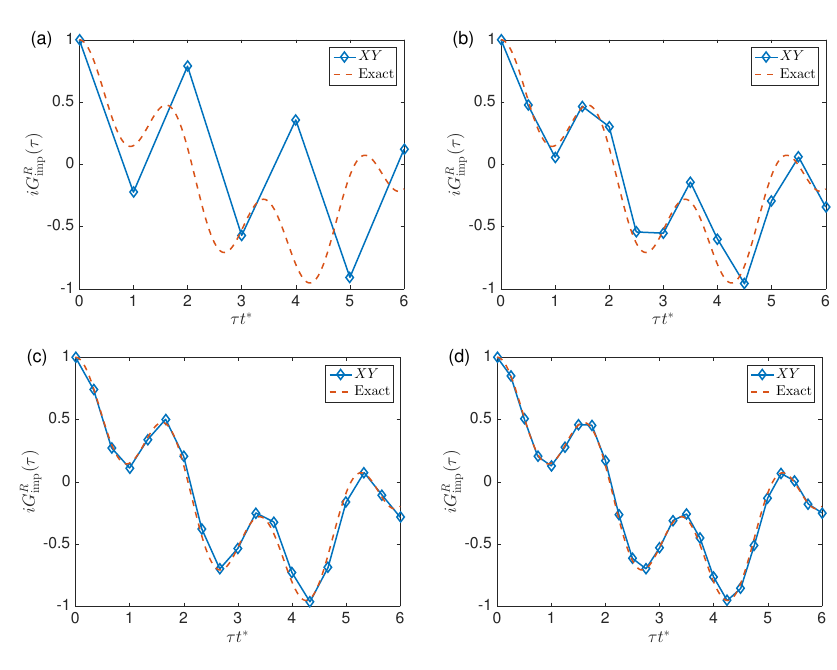}}
\caption{\csentence{Impurity Green function in the time domain.} The retarded impurity Green function $iG^R_{\rm imp}(\tau)$ obtained with (a) 6, (b) 12, (c) 18, and (d) 24 Trotter steps up to time $\tau=6/t^*$ using the $XY$ method (blue diamonds). Comparison is given to the exact Green function (red dashed line). We set $U=4t^*$ and $V=t^*$.}
\label{fig:grimp}
\end{figure}

As shown in Section~\ref{sec:2siteDMFT}, the main object of interest is the retarded impurity Green function. One possibility to measure $iG^R_{\rm imp}(\tau)$ is single-qubit interferometry (see the Appendix for details), which raises the total number of qubits in the experimental arrangement to five. In Fig.~\ref{fig:grimp} we plot the impurity Green function obtained from evolving the state with $XY$ gates compared to exact evolution of the two-site SIAM for different $N$. We see that the Green function from the $XY$ approach starts to follow the curve of the exact Green function better for increasing $N$. In our subsequent analysis, we use $N=24$ up to $\tau=6/t^*$ to study what two-site DMFT physics can be captured with the digital approach.

\begin{figure}
\centerline{\includegraphics[scale=1.45]{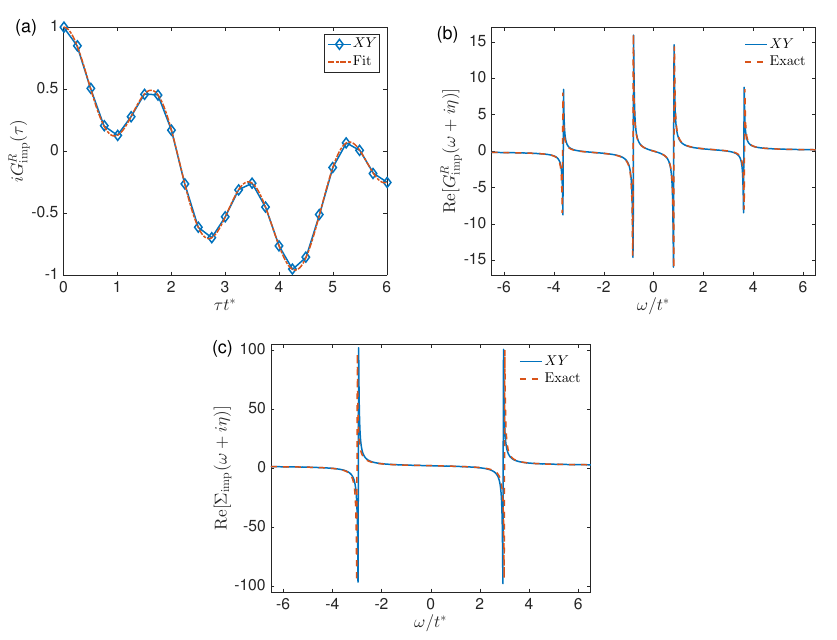}}
\caption{\csentence{The retarded impurity Green function and self-energy in the frequency domain}. (a) The residues and poles of the Green function can be obtained from a fit of the form in Eq.~\eqref{eq:GRcos} (red dashed line) to the $G^R_{\rm imp}(\tau)$ data from the $XY$ method with 24 Trotter steps (blue diamonds). (b) The real part of the impurity Green function, ${\rm Re}\left[ G^R_{\rm imp}(\omega+i\eta) \right]$ (blue line), with residues and poles obtained from the fit from (a), compared to the exact Green function (red dashed line). (c) Same as in (b), but for the self-energy ${\rm Re}\left[ \Sigma_{\rm imp}(\omega+i\eta) \right]$. We set $U=4t^*$ and $V=t^*$. In (b) and (c), we have broadened the peaks with $\eta=0.01$ for clarity.}
\label{fig:gself}
\end{figure}

To obtain the impurity Green function in the frequency domain, we first consider some known and general analytic properties of the retarded Green function in Eq.~\eqref{eq:Gtime}. This Green function can be written as a sum of the particle and hole contributions as
\begin{align}\label{eq:auxGR}
iG^R_{\rm imp}(\tau)=\theta(\tau)\sum_j \left( \big| \langle j | \hat{c}^{\dagger}_{1\sigma} | GS \rangle \big|^2 e^{-i\omega_j t}  +  \big| \langle j | \hat{c}_{1\sigma} | GS \rangle \big|^2 e^{i\omega_j t}  \right),
\end{align}
where $|j\rangle$ is an eigenstate of $\hat{H}_{\rm SIAM}$ with eigenenergy $E_j$, and $\omega_j=E_j-E_{GS}$. In two-site DMFT, the interacting Green function is a four-pole function~\cite{potthoff2001two}, which limits the number of terms in the above summation to four. Moreover, in the presence of particle-hole symmetry, we have $\big| \langle j | \hat{c}^{\dagger}_{1\sigma} | GS \rangle \big|^2= \big| \langle j | \hat{c}_{1\sigma} | GS \rangle \big|^2$, and Eq.~\eqref{eq:auxGR} can be written as
\begin{align}\label{eq:GRcos}
iG^R_{\rm imp}(\tau)=2\left[ \alpha_1 \cos(\omega_1\tau) +  \alpha_2 \cos(\omega_2 \tau) \right]\theta(\tau),
\end{align}
where $\alpha_j=\left| \langle j | \hat{c}^{\dagger}_{1\sigma} | GS \rangle \right|^2$. Thus, to obtain the impurity Green function in the frequency domain as
\begin{align}
G^R_{\rm imp}(\omega+i\eta)=&\alpha_1 \left( \frac{1}{\omega+i\eta-\omega_1} + \frac{1}{\omega+i\eta+\omega_1}  \right)\nonumber \\&+\alpha_2 \left( \frac{1}{\omega+i\eta-\omega_2} + \frac{1}{\omega+i\eta+\omega_2}  \right),
\end{align}
we need to extract the unknown residues $\alpha_j$ and poles $\omega_j$ by fitting an expression of the form  in Eq.~\eqref{eq:GRcos} to the measurement data of $iG^R_{\rm imp}(\tau)$, as shown in Fig.~\ref{fig:gself}a. This method to determine $\alpha_j$ and $\omega_j$ is far more reliable and requires fewer time steps than numerically Fourier-transforming the $iG^R_{\rm imp}(\tau)$ data. It can also be readily generalised to larger systems by including more terms in the sum in Eq.~\eqref{eq:auxGR}. Figure~\ref{fig:gself}b shows the real part of the impurity Green function in the frequency domain, ${\rm Re}\left[ G^R_{\rm imp}(\omega+i\eta) \right]$, with residues and poles obtained from the fit in Fig.~\ref{fig:gself}a, while in Fig.~\ref{fig:gself}c we plot the real part of the impurity self-energy, ${\rm Re}\left[ \Sigma_{\rm imp}(\omega+i\eta) \right]$, obtained utilizing the Dyson equation~\eqref{eq:dyson}. We clearly see the four-pole structure of the Green function, while the self-energy has two poles. The results are in excellent agreement with the exact solution of the two-site SIAM, with the poles of the self-energy using fitted $\alpha_j$ and $\omega_j$ differing from the exact solution by 2\%.

\begin{figure}[t!]
\centerline{\includegraphics[scale=1.45]{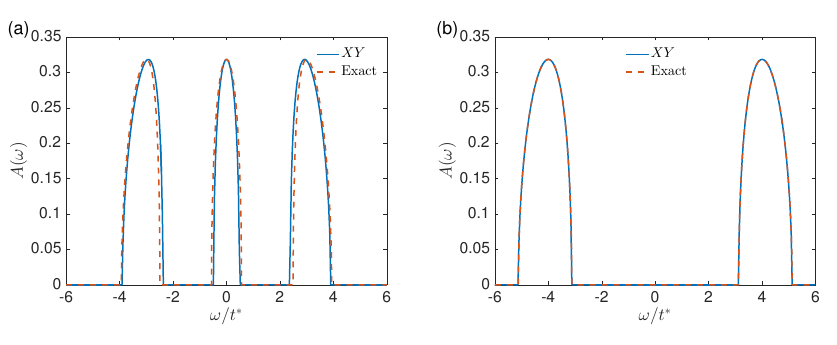}}
\caption{\csentence{Spectral functions in the metallic and insulating phases.} Spectral functions obtained with the $XY$ method with 24 Trotter steps (blue line) and exact solution of the two-site SIAM (red dashed line). The parameters of the two-site SIAM are iterated to self-consistency with (a) $U=5t^*$ and (b) $U=8t^*$.}
\label{fig:spec}
\end{figure}

\begin{figure}[h!]
\centerline{
\includegraphics[scale=0.39]{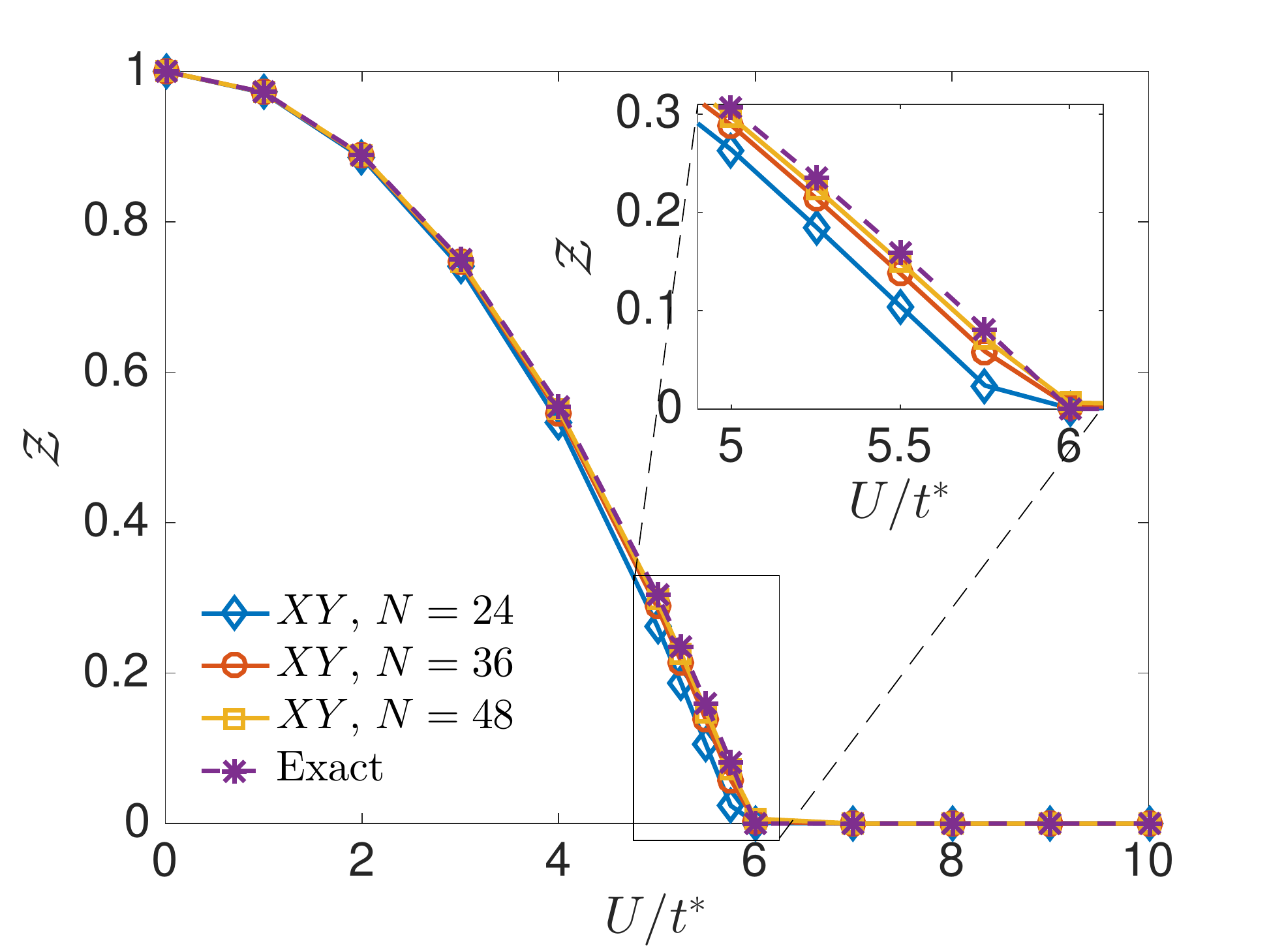}
}
\caption{\csentence{Quasiparticle weight as a function of interaction $U$.} Self-consistent quasiparticle weight $\mathcal{Z}$ obtained from the $XY$ method with 24 (blue diamonds), 36 (red circles), and 48 Trotter steps (yellow squares), compared to the exact solution of the two-site SIAM (purple stars). Inset: Same plot zoomed into the region around the critical interaction, $U_c=6t^*$.}
\label{fig:zvsu}
\end{figure}

Once we have obtained the impurity Green function, and thus the impurity self-energy, we proceed according to the two-site DMFT protocol in Section~\ref{sec:2siteDMFT} until self-consistency has been reached. In DMFT we are interested in the local lattice spectral function $A_{{\rm latt},jj}(\omega)$ which, at self-consistency, is given by the impurity spectral function $A_{\rm imp}(\omega)$. In the paramagnetic phase of the infinite-dimensional Hubbard model, the spectral function has a three peak structure with an upper and a lower Hubbard band, corresponding to empty and doubly occupied sites, respectively, and a quasiparticle peak with integrated spectral weight $\mathcal{Z}$ between the bands~\cite{georges1996dynamical}. In two-site DMFT, since the self-energy has two poles, this three peak structure can be qualitatively reproduced with the spectral function~\cite{potthoff2001two}
\begin{align}\label{eq:spectral}
A(\omega)=\rho_0\left[\omega+\mu-\Sigma_{\rm imp}(\omega)\right],
\end{align}
where $\rho_0$ is the non-interacting density of states of the Bethe lattice. Figure~\ref{fig:spec} shows the spectral function in Eq.~\eqref{eq:spectral} where the impurity self-energy has been obtained both from the $XY$ method and from exact numerics of the two-site SIAM using the interactions $U=5t^*$ and $U=8t^*$. We notice that for $U=5t^*$, the Hubbard bands from the $XY$ method are slightly dislocated and the quasiparticle peak is slightly narrower compared with the exact solution of the two-site SIAM, but the agreement is still very good. The overall shape of the spectral function from the $XY$ method is unchanged compared to the exact case. This underestimation of the width of the quasiparticle peak stems from the fact that the fitting procedure in Fig.~\ref{fig:gself}a causes the negative of the derivative of the self-energy in the $XY$ method to be a bit larger than the exact value from the two-site SIAM, i.e., $-\frac{d{\rm Re}[\Sigma^{XY}_{\rm imp}(\omega+i\eta)]}{d\omega}\Big|_{\omega=0} \gtrsim -\frac{d{\rm Re}[\Sigma^{\rm exact}_{\rm imp}(\omega+i\eta)]}{d\omega}\Big|_{\omega=0}$, which leads to $\mathcal{Z}$ in Eq.~\eqref{eq:Z} from the $XY$ method to be slightly smaller than in the exact solution of the two-site SIAM, i.e., $\mathcal{Z}^{XY} \lesssim \mathcal{Z}^{\rm exact}$.  For $U=8t^*$, the two spectral functions agree with maximum relative error of $10^{-8}$, since in this case $V=0$ is found to be the self-consistent solution, whence the Trotterized evolution operator in Eq.~\eqref{eq:TrotterU} matches full evolution operator of the two-site SIAM, and thus there is no Trotter error. We observe that in Fig.~\ref{fig:spec} the central quasiparticle peak vanishes, which is characteristic of insulating behaviour. See Ref.~\cite{potthoff2001two} for a discussion of the artifacts of the spectral functions in two-site DMFT compared to full DMFT.

To study the transition between the two types of spectral functions in Fig.~\ref{fig:spec}, we plot in Fig.~\ref{fig:zvsu} the self-consistent quasiparticle weight $\mathcal{Z}$ obtained from the $XY$ method as a function of the interaction $U$ for different Trotter steps $N$. We also show $\mathcal{Z}$ from the exact solution  of the two-site SIAM for comparison. We see that the digital approach captures the correct trend of the curve, but in the metallic side underestimates to a small degree the values of $\mathcal{Z}$ for interactions close to $U=U_c=6t^*$, which is the critical interaction for Mott transition in two-site DMFT at half-filling~\cite{potthoff2001two}. These results are consistent with the spectral functions in Fig.~\ref{fig:spec}. The underestimation of $\mathcal{Z}$ can be diminished by increasing $N$, as shown in Fig.~\ref{fig:zvsu}. It is noteworthy to mention that two-site DMFT overestimates the quasiparticle weight compared to full DMFT for interactions $U<U_c$, as demonstrated in Ref.~\cite{potthoff2001two}.  Above $U_c$, we find $\mathcal{Z}=0$ to be the self-consistent solution, corresponding to the insulating phase. 

\section{Summary}\label{sec:summ}
We have proposed a quantum algorithm for two-site DMFT to be run on a small digital quantum simulator with a classical feedback loop, allowing the qualitative description of the infinite-dimensional Hubbard model in the thermodynamic limit. We have considered two alternative quantum gate decompositions consistent with state-of-the-art technology in superconducting circuits for the time-evolution operator. We found that an increasing number of Trotter steps improves the fidelity of our digital scheme to qualitatively describe the Mott transition. Our work therefore provides an interesting application for small-scale quantum devices. It also paves the way for more accurate quantum simulations of strongly correlated fermions in various lattice geometries, which are relevant to novel quantum materials, when the general self-consistency condition and larger number of qubits are used.

\section*{Appendix: Single-qubit interferometry for the impurity Green function}\label{sec:sqinter}
Here, we present a measurement scheme for the retarded impurity Green function.
\subsection*{Definitions}
The retarded zero temperature impurity Green function in the time domain can be written as
\begin{align}
G^R_{\rm imp}(\tau)=\theta(\tau) \left[ G^{>}_{\rm imp}(\tau)-G^{<}_{\rm imp}(\tau) \right],
\end{align}
where the ``greater'' and ``lesser'' Green functions are given by
\begin{align}\label{eq:greater}
G^{>}_{\rm imp}(\tau)=-i\langle \hat{c}_{1\sigma}(\tau) \hat{c}_{1\sigma}^{\dagger}(0) \rangle,
\end{align}
\begin{align}\label{eq:lesser}
G^{<}_{\rm imp}(\tau)=i\langle \hat{c}^{\dagger}_{1\sigma}(0) \hat{c}_{1\sigma}(\tau) \rangle,
\end{align}
respectively. The average is computed in the ground-state $|GS\rangle$ of the two-site SIAM in Eq.~\eqref{eq:HSIAM}.
Here, $\sigma$ can be either $\downarrow$ or $\uparrow$ since we are considering a spin-symmetric case (i.e., $G^R_{\downarrow}=G^R_{\uparrow}$), and the $\hat{c}$-operators are given in the Heisenberg picture with respect to $\hat{H}_{\mathrm{SIAM}}$, i.e.,
\begin{align}
\hat{c}_{1\sigma}(\tau)=\hat{U}^{\dagger}(\tau)\hat{c}_{1\sigma}\hat{U}(\tau)=e^{i\tau\hat{H}_{\rm SIAM}}  \hat{c}_{1\sigma} e^{-i\tau\hat{H}_{\rm SIAM}}.
\end{align}

One possibility to measure the impurity Green function $G^R_{\rm imp}(\tau)$ is to use a single-qubit Ramsey interferometer~\cite{dorner2013extracting} which was used in Ref.~\cite{kreula2015coprocessor} in the more general non-equilibrium case. To this end, we introduce an ancilla qubit in addition to the `system' qubits, raising the total number of qubits needed to implement the two-site DMFT scheme to five.

\subsection*{Jordan--Wigner transformation}
The greater and lesser components, $G^{>}_{\rm imp}(\tau)$ and $G^{<}_{\rm imp}(\tau)$, must be written in terms of spin operators by again mapping the $\hat{c}_{1\sigma}$ and $\hat{c}^{\dagger}_{1\sigma}$ operators onto Pauli operators via the Jordan--Wigner transformation. For concreteness, we focus on the case $\sigma = \downarrow$. We obtain
\begin{align}
G^{>}_{\rm imp}(\tau)=-\frac{i}{4}\Big( &\langle \hat{U}^{\dagger}(\tau)\hat{\sigma}^x_1 \hat{U}(\tau) \hat{\sigma}^x_1 \rangle + i\langle  \hat{U}^{\dagger}(\tau)\hat{\sigma}^x_1 \hat{U}(\tau) \hat{\sigma}^y_1 \rangle  -i\langle \hat{U}^{\dagger}(\tau)\hat{\sigma}^y_1 \hat{U}(\tau) \hat{\sigma}^x_1 \rangle \nonumber \\ &+ \langle  \hat{U}^{\dagger}(\tau)\hat{\sigma}^y_1 \hat{U}(\tau) \hat{\sigma}^y_1 \rangle \Big),
\end{align}
and
\begin{align}
G^{<}_{\rm imp}(\tau)=\frac{i}{4}\Big( &\langle  \hat{\sigma}^x_1 \hat{U}^{\dagger}(\tau)  \hat{\sigma}^x_1\hat{U}(\tau)\rangle - i\langle  \hat{\sigma}^x_1 \hat{U}^{\dagger}(\tau) \hat{\sigma}^y_1\hat{U}(\tau) \rangle   +i\langle  \hat{\sigma}^y_1 \hat{U}^{\dagger}(\tau) \hat{\sigma}^x_1 \hat{U}(\tau)\rangle \nonumber \\ & + \langle  \hat{\sigma}^y_1 \hat{U}^{\dagger}(\tau) \hat{\sigma}^y_1\hat{U}(\tau) \rangle \Big).
\end{align}
\begin{figure}[ht!]
\centerline{
\includegraphics[scale=1]{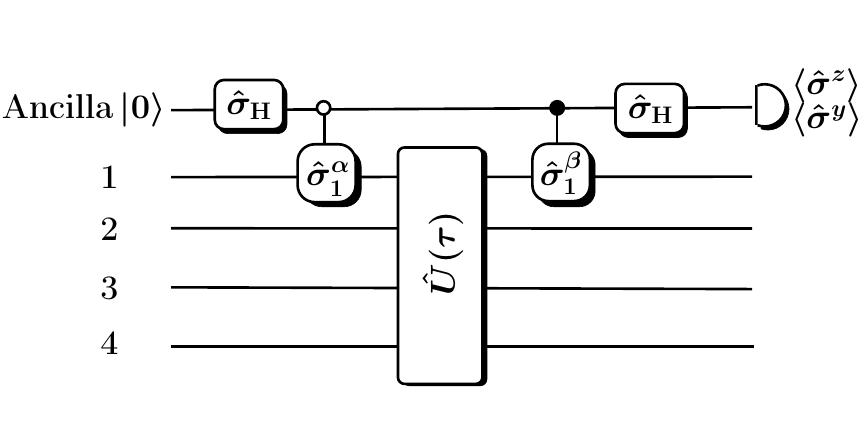}
}
\caption{\csentence{Quantum network to measure the $\langle GS |  \hat{U}^{\dagger}(\tau) \hat{\sigma}^{\alpha}_1 \hat{U}(\tau) \hat{\sigma}^{\beta}_1 |GS\rangle$ contribution to the Green function $G^R_{\rm imp}(\tau)$.} The time-evolution operator $\hat{U}(\tau)$ is composed of a set of quantum gates according to the main text.}
\label{fig:twositeramsey}
\end{figure}
\subsection*{Measurement protocol}
Each of the terms of the form $\langle   \hat{U}^{\dagger}(\tau) \hat{\sigma}^{\alpha}_1 \hat{U}(\tau) \hat{\sigma}^{\beta}_1 \rangle$, where $\alpha, \beta \in \{x, y\}$, can be measured in the interferometer. This can be seen as follows. We denote the state of the system qubits by $\hat{\rho}_{\rm sys}=|GS\rangle \langle GS|$, where $|GS\rangle$ is the ground-state of the system. We initialize the ancilla qubit in the state $|0\rangle$, yielding the total density operator $\hat{\rho}_{\rm tot}=|0\rangle \langle 0 | \otimes \hat{\rho}_{\rm sys}$. The total system then undergoes the following evolution:
\begin{itemize}
\item[1.] At time $t=0$, a Hadamard gate $\hat{\sigma}_H=\frac{1}{\sqrt{2}}\left(\hat{\sigma}^z+\hat{\sigma}^x \right)$ is applied on the ancilla qubit, creating the superposition $|0\rangle_{\rm ancilla} \rightarrow \left(|0\rangle_{\rm ancilla} + |1\rangle_{\rm ancilla} \right)/\sqrt{2}$.
\item[2.] A Controlled-Pauli gate $\hat{\sigma}^{\alpha}_1$ is applied on the impurity qubit 1 if the ancilla qubit has state $|0\rangle$.
\item[3.] The system qubits undergo time evolution according to the unitary $\hat{U}(\tau)$ which is decomposed into quantum gates.
\item[4.] Another controlled Pauli gate $\hat{\sigma}^{\beta}_1$ is applied on the impurity qubit 1 if the ancilla qubit has state $|1\rangle$.
\item[5.] Another Hadamard gate is applied on the ancilla qubit.
\end{itemize}

Denoting the total unitary in steps 2-4 by $\hat{T}$, the state of the ancilla qubit after this evolution is given by
\begin{align}
\hat{\rho}_{\rm ancilla}&=\tr_{\rm sys}\left[ \hat{\sigma}_H \hat{T} \hat{\sigma}_H \hat{\rho}_{\rm tot}  \hat{\sigma}_H \hat{T}^{\dagger} \hat{\sigma}_H \right] \nonumber \\
&=\frac{1+\mathrm{Re}[F(\tau)]}{2}|0\rangle \langle 0 | - i\frac{ \mathrm{Im}[F(\tau)]}{2}|0 \rangle \langle 1 |  + i\frac{ \mathrm{Im}[F(\tau)]}{2}|1 \rangle \langle 0 |\nonumber \\& + \frac{1-\mathrm{Re}[F(\tau)]}{2}|1 \rangle \langle 1 |,
\end{align}
where $F(\tau)=\tr_{\mathrm{sys}}\left[ \hat{T}_{1}^{\dagger}(\tau) \hat{T}_{0}(\tau) \hat{\rho}_{\mathrm{sys}} \right]$. We have denoted the controlled unitaries as $\hat{T}_{1}(\tau)=\hat{\sigma}^{\alpha}_1 \hat{U}(\tau)$ and $\hat{T}_{0}(\tau)= \hat{U}(\tau)\hat{\sigma}^{\beta}_1$. Note that since the same $\hat{U}(\tau)$ appears in both unitaries, only the Pauli gates $\hat{\sigma}^{\alpha / \beta}_1$ need to be controlled, as described above. Note also that $F(\tau)=\langle \hat{U}^{\dagger}(\tau) \hat{\sigma}^{\alpha}_1 \hat{U}(\tau) \hat{\sigma}^{\beta}_1 \rangle$. We can rewrite the state of the ancilla qubit as
\begin{align}
\hat{\rho}_{\mathrm{ancilla}}=\frac{1}{2}\left(\hat{I} + \mathrm{Re}[F(\tau)]\hat{\sigma}_z +\mathrm{Im}[F(\tau)] \hat{\sigma}_y \right),
\end{align}
whence
$
\tr_{\mathrm{ancilla}}\left[\hat{\rho}_{\mathrm{ancilla}} \hat{\sigma}^z \right]=\mathrm{Re}[F(\tau)],
$
and
$
\tr_{\mathrm{ancilla}}\left[\hat{\rho}_{\mathrm{ancilla}} \hat{\sigma}^y \right]=\mathrm{Im}[F(\tau)].
$
Thus, repeated measurements of the $\hat{\sigma}^z$ and $\hat{\sigma}^y$ components of the ancilla qubit yield the real and imaginary parts of the term $\langle  \hat{U}^{\dagger}(\tau) \hat{\sigma}^{\alpha}_1 \hat{U}(\tau) \hat{\sigma}^{\beta}_1 \rangle$. See Fig.~\ref{fig:twositeramsey} for the quantum network of the scheme.


\begin{backmatter}

\section*{Competing interests}
  The authors declare that they have no competing interests.

\section*{Author's contributions}
JMK conceived the project and performed the numerical simulations. LG-\'{A}, LL and ES worked out the superconducting circuit implementation. All authors contributed to interpreting the results and writing of the manuscript.

\section*{Acknowledgements}
  We acknowledge Anna-Maija Uimonen as well as Ian Walmsley and his group members for useful discussions. JMK acknowledges financial support from Christ Church, Oxford and the Osk Huttunen Foundation. LG-\'{A}, LL and ES acknowledge support from a UPV/EHU PhD grant, Spanish MINECO FIS2015-69983-P, UPV/EHU UFI 11/55 and Project EHUA14/04, and Ram\'on y Cajal Grant RYC-2012-11391. DJ was supported by the EPSRC National Quantum Technology Hub in Networked Quantum Information Processing (NQIT) EP/M013243/1.




\newcommand{\BMCxmlcomment}[1]{}

\BMCxmlcomment{

<refgrp>

<bibl id="B1">
  <title><p>Simulating physics with computers</p></title>
  <aug>
    <au><snm>Feynman</snm><fnm>RP</fnm></au>
  </aug>
  <source>Int. J. Theor. Phys.</source>
  <publisher>World Scientific</publisher>
  <pubdate>1982</pubdate>
  <volume>21</volume>
  <issue>6/7</issue>
  <fpage>467</fpage>
  <lpage>-488</lpage>
</bibl>

<bibl id="B2">
  <title><p>Quantum simulators</p></title>
  <aug>
    <au><snm>Buluta</snm><fnm>I</fnm></au>
    <au><snm>Nori</snm><fnm>F</fnm></au>
  </aug>
  <source>Science</source>
  <publisher>American Association for the Advancement of Science</publisher>
  <pubdate>2009</pubdate>
  <volume>326</volume>
  <issue>5949</issue>
  <fpage>108</fpage>
  <lpage>-111</lpage>
</bibl>

<bibl id="B3">
  <title><p>Goals and opportunities in quantum simulation</p></title>
  <aug>
    <au><snm>Cirac</snm><fnm>JI</fnm></au>
    <au><snm>Zoller</snm><fnm>P</fnm></au>
  </aug>
  <source>Nature Phys.</source>
  <publisher>Nature Publishing Group</publisher>
  <pubdate>2012</pubdate>
  <volume>8</volume>
  <issue>4</issue>
  <fpage>264</fpage>
  <lpage>-266</lpage>
</bibl>

<bibl id="B4">
  <title><p>What is a quantum simulator?</p></title>
  <aug>
    <au><snm>Johnson</snm><fnm>TH</fnm></au>
    <au><snm>Clark</snm><fnm>SR</fnm></au>
    <au><snm>Jaksch</snm><fnm>D</fnm></au>
  </aug>
  <source>EPJ Quantum Technology</source>
  <publisher>Springer</publisher>
  <pubdate>2014</pubdate>
  <volume>1</volume>
  <issue>1</issue>
  <fpage>1</fpage>
  <lpage>-12</lpage>
</bibl>

<bibl id="B5">
  <title><p>The rise of quantum materials</p></title>
  <source>Nature Phys.</source>
  <pubdate>2016</pubdate>
  <volume>12</volume>
  <fpage>105</fpage>
</bibl>

<bibl id="B6">
  <title><p>Metal-insulator transition</p></title>
  <aug>
    <au><snm>Mott</snm><fnm>N F</fnm></au>
  </aug>
  <source>Rev. Mod. Phys.</source>
  <publisher>APS</publisher>
  <pubdate>1968</pubdate>
  <volume>40</volume>
  <issue>4</issue>
  <fpage>677</fpage>
</bibl>

<bibl id="B7">
  <title><p>Metal-insulator transitions</p></title>
  <aug>
    <au><snm>Imada</snm><fnm>M</fnm></au>
    <au><snm>Fujimori</snm><fnm>A</fnm></au>
    <au><snm>Tokura</snm><fnm>Y</fnm></au>
  </aug>
  <source>Rev. Mod. Phys.</source>
  <publisher>APS</publisher>
  <pubdate>1998</pubdate>
  <volume>70</volume>
  <issue>4</issue>
  <fpage>1039</fpage>
</bibl>

<bibl id="B8">
  <title><p>Colossal magnetoresistance</p></title>
  <aug>
    <au><snm>Ramirez</snm><fnm>A P</fnm></au>
  </aug>
  <source>Journal of Physics: Condensed Matter</source>
  <publisher>IOP Publishing</publisher>
  <pubdate>1997</pubdate>
  <volume>9</volume>
  <issue>39</issue>
  <fpage>8171</fpage>
</bibl>

<bibl id="B9">
  <title><p>Doping a {M}ott insulator: {P}hysics of high-temperature
  superconductivity</p></title>
  <aug>
    <au><snm>Lee</snm><fnm>PA</fnm></au>
    <au><snm>Nagaosa</snm><fnm>N</fnm></au>
    <au><snm>Wen</snm><fnm>XG</fnm></au>
  </aug>
  <source>Rev. Mod. Phys.</source>
  <publisher>APS</publisher>
  <pubdate>2006</pubdate>
  <volume>78</volume>
  <issue>1</issue>
  <fpage>17</fpage>
</bibl>

<bibl id="B10">
  <title><p>Handbook of {H}igh-{T}emperature {S}uperconductivity</p></title>
  <publisher>New York: Springer-Verlag</publisher>
  <editor>Schrieffer, J Robert</editor>
  <edition>1</edition>
  <pubdate>2007</pubdate>
</bibl>

<bibl id="B11">
  <title><p>Quantum {M}onte {C}arlo simulations of solids</p></title>
  <aug>
    <au><snm>Foulkes</snm><fnm>W M C</fnm></au>
    <au><snm>Mitas</snm><fnm>L</fnm></au>
    <au><snm>Needs</snm><fnm>R J</fnm></au>
    <au><snm>Rajagopal</snm><fnm>G</fnm></au>
  </aug>
  <source>Rev. Mod. Phys.</source>
  <publisher>APS</publisher>
  <pubdate>2001</pubdate>
  <volume>73</volume>
  <issue>1</issue>
  <fpage>33</fpage>
</bibl>

<bibl id="B12">
  <title><p>Continuous-time quantum {M}onte {C}arlo method for
  fermions</p></title>
  <aug>
    <au><snm>Rubtsov</snm><fnm>AN</fnm></au>
    <au><snm>Savkin</snm><fnm>VV</fnm></au>
    <au><snm>Lichtenstein</snm><fnm>AI</fnm></au>
  </aug>
  <source>Phys. Rev. B</source>
  <publisher>APS</publisher>
  <pubdate>2005</pubdate>
  <volume>72</volume>
  <issue>3</issue>
  <fpage>035122</fpage>
</bibl>

<bibl id="B13">
  <title><p>Computational complexity and fundamental limitations to fermionic
  quantum {M}onte {C}arlo simulations</p></title>
  <aug>
    <au><snm>Troyer</snm><fnm>M</fnm></au>
    <au><snm>Wiese</snm><fnm>UJ</fnm></au>
  </aug>
  <source>Phys. Rev. Lett.</source>
  <publisher>APS</publisher>
  <pubdate>2005</pubdate>
  <volume>94</volume>
  <issue>17</issue>
  <fpage>170201</fpage>
</bibl>

<bibl id="B14">
  <title><p>Efficient classical simulation of slightly entangled quantum
  computations</p></title>
  <aug>
    <au><snm>Vidal</snm><fnm>G.</fnm></au>
  </aug>
  <source>Phys. Rev. Lett.</source>
  <publisher>APS</publisher>
  <pubdate>2003</pubdate>
  <volume>91</volume>
  <issue>14</issue>
  <fpage>147902</fpage>
</bibl>

<bibl id="B15">
  <title><p>Efficient simulation of one-dimensional quantum many-body
  systems</p></title>
  <aug>
    <au><snm>Vidal</snm><fnm>G.</fnm></au>
  </aug>
  <source>Phys. Rev. Lett.</source>
  <publisher>APS</publisher>
  <pubdate>2004</pubdate>
  <volume>93</volume>
  <issue>4</issue>
  <fpage>40502</fpage>
</bibl>

<bibl id="B16">
  <title><p>Matrix product states, projected entangled pair states, and
  variational renormalization group methods for quantum spin
  systems</p></title>
  <aug>
    <au><snm>Verstraete</snm><fnm>F.</fnm></au>
    <au><snm>Murg</snm><fnm>V.</fnm></au>
    <au><snm>Cirac</snm><fnm>J. I.</fnm></au>
  </aug>
  <source>Adv. Phys.</source>
  <publisher>Taylor \& Francis</publisher>
  <pubdate>2008</pubdate>
  <volume>57</volume>
  <issue>2</issue>
  <fpage>143</fpage>
  <lpage>-224</lpage>
</bibl>

<bibl id="B17">
  <title><p>Renormalization and tensor product states in spin chains and
  lattices</p></title>
  <aug>
    <au><snm>Cirac</snm><fnm>JI</fnm></au>
    <au><snm>Verstraete</snm><fnm>F</fnm></au>
  </aug>
  <source>Journal of Physics A: Mathematical and Theoretical</source>
  <publisher>IOP Publishing</publisher>
  <pubdate>2009</pubdate>
  <volume>42</volume>
  <issue>50</issue>
  <fpage>504004</fpage>
</bibl>

<bibl id="B18">
  <title><p>The density-matrix renormalization group in the age of matrix
  product states</p></title>
  <aug>
    <au><snm>Schollw{\"o}ck</snm><fnm>U</fnm></au>
  </aug>
  <source>Ann. Phys.</source>
  <publisher>Elsevier</publisher>
  <pubdate>2011</pubdate>
  <volume>326</volume>
  <issue>1</issue>
  <fpage>96</fpage>
  <lpage>-192</lpage>
</bibl>

<bibl id="B19">
  <title><p>Dynamical mean-field theory of strongly correlated fermion systems
  and the limit of infinite dimensions</p></title>
  <aug>
    <au><snm>Georges</snm><fnm>A</fnm></au>
    <au><snm>Kotliar</snm><fnm>G</fnm></au>
    <au><snm>Krauth</snm><fnm>W</fnm></au>
    <au><snm>Rozenberg</snm><fnm>MJ</fnm></au>
  </aug>
  <source>Rev. Mod. Phys.</source>
  <publisher>APS</publisher>
  <pubdate>1996</pubdate>
  <volume>68</volume>
  <issue>1</issue>
  <fpage>13</fpage>
</bibl>

<bibl id="B20">
  <title><p>Electron correlations in narrow energy bands</p></title>
  <aug>
    <au><snm>Hubbard</snm><fnm>J.</fnm></au>
  </aug>
  <source>Proc. Roy. Soc. London Ser. A</source>
  <publisher>The Royal Society</publisher>
  <pubdate>1963</pubdate>
  <volume>276</volume>
  <issue>1365</issue>
  <fpage>238</fpage>
  <lpage>-257</lpage>
</bibl>

<bibl id="B21">
  <title><p>Quantum simulations with ultracold quantum gases</p></title>
  <aug>
    <au><snm>Bloch</snm><fnm>I</fnm></au>
    <au><snm>Dalibard</snm><fnm>J</fnm></au>
    <au><snm>Nascimb{\`e}ne</snm><fnm>S</fnm></au>
  </aug>
  <source>Nature Phys.</source>
  <publisher>Nature Publishing Group</publisher>
  <pubdate>2012</pubdate>
  <volume>8</volume>
  <issue>4</issue>
  <fpage>267</fpage>
  <lpage>-276</lpage>
</bibl>

<bibl id="B22">
  <title><p>Universal quantum simulators</p></title>
  <aug>
    <au><snm>Lloyd</snm><fnm>S</fnm></au>
  </aug>
  <source>Science</source>
  <publisher>American Association for the Advancement of Science</publisher>
  <pubdate>1996</pubdate>
  <volume>273</volume>
  <fpage>1073</fpage>
  <lpage>-1078</lpage>
</bibl>

<bibl id="B23">
  <title><p>Cavity quantum electrodynamics for superconducting electrical
  circuits: An architecture for quantum computation</p></title>
  <aug>
    <au><snm>Blais</snm><fnm>A</fnm></au>
    <au><snm>Huang</snm><fnm>RS</fnm></au>
    <au><snm>Wallraff</snm><fnm>A</fnm></au>
    <au><snm>Girvin</snm><fnm>S M</fnm></au>
    <au><snm>Schoelkopf</snm><fnm>RJ</fnm></au>
  </aug>
  <source>Phys. Rev. A</source>
  <publisher>APS</publisher>
  <pubdate>2004</pubdate>
  <volume>69</volume>
  <issue>6</issue>
  <fpage>062320</fpage>
</bibl>

<bibl id="B24">
  <title><p>On-chip quantum simulation with superconducting
  circuits</p></title>
  <aug>
    <au><snm>Houck</snm><fnm>AA</fnm></au>
    <au><snm>T{\"u}reci</snm><fnm>HE</fnm></au>
    <au><snm>Koch</snm><fnm>J</fnm></au>
  </aug>
  <source>Nature Phys.</source>
  <publisher>Nature Publishing Group</publisher>
  <pubdate>2012</pubdate>
  <volume>8</volume>
  <issue>4</issue>
  <fpage>292</fpage>
  <lpage>-299</lpage>
</bibl>

<bibl id="B25">
  <title><p>Digital quantum simulation of fermionic models with a
  superconducting circuit</p></title>
  <aug>
    <au><snm>Barends</snm><fnm>R</fnm></au>
    <au><cnm>others</cnm></au>
  </aug>
  <source>Nature Comm.</source>
  <pubdate>2015</pubdate>
  <volume>6</volume>
  <fpage>7654</fpage>
</bibl>

<bibl id="B26">
  <title><p>Digitized adiabatic quantum computing with a superconducting
  circuit</p></title>
  <aug>
    <au><snm>Barends</snm><fnm>R</fnm></au>
    <au><cnm>others</cnm></au>
  </aug>
  <source>Nature</source>
  <pubdate>2016</pubdate>
  <volume>534</volume>
  <fpage>222</fpage>
</bibl>

<bibl id="B27">
  <title><p>Two-site dynamical mean-field theory</p></title>
  <aug>
    <au><snm>Potthoff</snm><fnm>M</fnm></au>
  </aug>
  <source>Phys. Rev. B</source>
  <publisher>APS</publisher>
  <pubdate>2001</pubdate>
  <volume>64</volume>
  <issue>16</issue>
  <fpage>165114</fpage>
</bibl>

<bibl id="B28">
  <title><p>IBM Quantum Computing</p></title>
  <source>\url{https://www.research.ibm.com/quantum/}</source>
  <note>{(Retrieved 14 June, 2016)}</note>
</bibl>

<bibl id="B29">
  <title><p>State preservation by repetitive error detection in a
  superconducting quantum circuit</p></title>
  <aug>
    <au><snm>Kelly</snm><fnm>J</fnm></au>
    <au><snm>Barends</snm><fnm>R</fnm></au>
    <au><snm>Fowler</snm><fnm>A G</fnm></au>
    <au><snm>Megrant</snm><fnm>A</fnm></au>
    <au><snm>Jeffrey</snm><fnm>E</fnm></au>
    <au><snm>White</snm><fnm>T C</fnm></au>
    <au><snm>Sank</snm><fnm>D</fnm></au>
    <au><snm>Mutus</snm><fnm>J Y</fnm></au>
    <au><snm>Campbell</snm><fnm>B</fnm></au>
    <au><snm>Chen</snm><fnm>Y</fnm></au>
    <au><cnm>others</cnm></au>
  </aug>
  <source>Nature</source>
  <publisher>Nature Publishing Group</publisher>
  <pubdate>2015</pubdate>
  <volume>519</volume>
  <issue>7541</issue>
  <fpage>66</fpage>
  <lpage>-69</lpage>
</bibl>

<bibl id="B30">
  <title><p>Universal digital quantum simulation with trapped ions</p></title>
  <aug>
    <au><snm>Lanyon</snm><fnm>B P</fnm></au>
    <au><snm>Hempel</snm><fnm>C</fnm></au>
    <au><snm>Nigg</snm><fnm>D</fnm></au>
    <au><snm>M{\"u}ller</snm><fnm>M</fnm></au>
    <au><snm>Gerritsma</snm><fnm>R</fnm></au>
    <au><snm>Z{\"a}hringer</snm><fnm>F</fnm></au>
    <au><snm>Schindler</snm><fnm>P</fnm></au>
    <au><snm>Barreiro</snm><fnm>J T</fnm></au>
    <au><snm>Rambach</snm><fnm>M</fnm></au>
    <au><snm>Kirchmair</snm><fnm>G</fnm></au>
    <au><cnm>others</cnm></au>
  </aug>
  <source>Science</source>
  <publisher>American Association for the Advancement of Science</publisher>
  <pubdate>2011</pubdate>
  <volume>334</volume>
  <issue>6052</issue>
  <fpage>57</fpage>
  <lpage>-61</lpage>
</bibl>

<bibl id="B31">
  <title><p>Real-time dynamics of lattice gauge theories with a few-qubit
  quantum computer</p></title>
  <aug>
    <au><snm>Martinez</snm><fnm>E. A.</fnm></au>
    <au><cnm>others</cnm></au>
  </aug>
  <source>arXiv:1605.04570</source>
  <pubdate>2016</pubdate>
</bibl>

<bibl id="B32">
  <title><p>A quantum coprocessor for accelerating simulations of
  non-equilibrium many body quantum dynamics</p></title>
  <aug>
    <au><snm>Kreula</snm><fnm>J. M.</fnm></au>
    <au><snm>Clark</snm><fnm>S. R.</fnm></au>
    <au><snm>Jaksch</snm><fnm>D.</fnm></au>
  </aug>
  <source>arXiv:1510.05703</source>
  <pubdate>2015</pubdate>
</bibl>

<bibl id="B33">
  <title><p>Hybrid quantum-classical approach to correlated
  materials</p></title>
  <aug>
    <au><snm>Bauer</snm><fnm>B</fnm></au>
    <au><snm>Wecker</snm><fnm>D</fnm></au>
    <au><snm>Millis</snm><fnm>AJ</fnm></au>
    <au><snm>Hastings</snm><fnm>MB</fnm></au>
    <au><snm>Troyer</snm><fnm>M.</fnm></au>
  </aug>
  <source>arXiv:1510.03859</source>
  <pubdate>2015</pubdate>
</bibl>

<bibl id="B34">
  <title><p>Fermionic models with superconducting circuits</p></title>
  <aug>
    <au><snm>Las Heras</snm><fnm>U</fnm></au>
    <au><snm>Garc{\'\i}a {\'A}lvarez</snm><fnm>L</fnm></au>
    <au><snm>Mezzacapo</snm><fnm>A</fnm></au>
    <au><snm>Solano</snm><fnm>E</fnm></au>
    <au><snm>Lamata</snm><fnm>L</fnm></au>
  </aug>
  <source>EPJ Quantum Technology</source>
  <publisher>Springer</publisher>
  <pubdate>2015</pubdate>
  <volume>2</volume>
  <issue>1</issue>
  <fpage>1</fpage>
  <lpage>-11</lpage>
</bibl>

<bibl id="B35">
  <title><p>Towards fault-tolerant quantum computing with trapped
  ions</p></title>
  <aug>
    <au><snm>Benhelm</snm><fnm>J</fnm></au>
    <au><snm>Kirchmair</snm><fnm>G</fnm></au>
    <au><snm>Roos</snm><fnm>CF</fnm></au>
    <au><snm>Blatt</snm><fnm>R</fnm></au>
  </aug>
  <source>Nature Phys.</source>
  <publisher>Nature Publishing Group</publisher>
  <pubdate>2008</pubdate>
  <volume>4</volume>
  <issue>6</issue>
  <fpage>463</fpage>
  <lpage>-466</lpage>
</bibl>

<bibl id="B36">
  <title><p>Quantum simulations with trapped ions</p></title>
  <aug>
    <au><snm>Blatt</snm><fnm>R</fnm></au>
    <au><snm>Roos</snm><fnm>C F</fnm></au>
  </aug>
  <source>Nature Phys.</source>
  <publisher>Nature Publishing Group</publisher>
  <pubdate>2012</pubdate>
  <volume>8</volume>
  <issue>4</issue>
  <fpage>277</fpage>
  <lpage>-284</lpage>
</bibl>

<bibl id="B37">
  <title><p>Quantum simulation of interacting fermion lattice models in trapped
  ions</p></title>
  <aug>
    <au><snm>Casanova</snm><fnm>J</fnm></au>
    <au><snm>Mezzacapo</snm><fnm>A</fnm></au>
    <au><snm>Lamata</snm><fnm>L</fnm></au>
    <au><snm>Solano</snm><fnm>E</fnm></au>
  </aug>
  <source>Phys. Rev. Lett.</source>
  <publisher>APS</publisher>
  <pubdate>2012</pubdate>
  <volume>108</volume>
  <issue>19</issue>
  <fpage>190502</fpage>
</bibl>

<bibl id="B38">
  <title><p>Superconducting quantum circuits at the surface code threshold for
  fault tolerance</p></title>
  <aug>
    <au><snm>Barends</snm><fnm>R</fnm></au>
    <au><snm>Kelly</snm><fnm>J</fnm></au>
    <au><snm>Megrant</snm><fnm>A</fnm></au>
    <au><snm>Veitia</snm><fnm>A</fnm></au>
    <au><snm>Sank</snm><fnm>D</fnm></au>
    <au><snm>Jeffrey</snm><fnm>E</fnm></au>
    <au><snm>White</snm><fnm>T C</fnm></au>
    <au><snm>Mutus</snm><fnm>J</fnm></au>
    <au><snm>Fowler</snm><fnm>A G</fnm></au>
    <au><snm>Campbell</snm><fnm>B</fnm></au>
    <au><cnm>others</cnm></au>
  </aug>
  <source>Nature</source>
  <publisher>Nature Publishing Group</publisher>
  <pubdate>2014</pubdate>
  <volume>508</volume>
  <issue>7497</issue>
  <fpage>500</fpage>
  <lpage>-503</lpage>
</bibl>

<bibl id="B39">
  <title><p>Digital Quantum Simulation of Spin Systems in Superconducting
  Circuits</p></title>
  <aug>
    <au><snm>Las Heras</snm><fnm>U.</fnm></au>
    <au><snm>Mezzacapo</snm><fnm>A.</fnm></au>
    <au><snm>Lamata</snm><fnm>L.</fnm></au>
    <au><snm>Filipp</snm><fnm>S.</fnm></au>
    <au><snm>Wallraff</snm><fnm>A.</fnm></au>
    <au><snm>Solano</snm><fnm>E.</fnm></au>
  </aug>
  <source>Phys. Rev. Lett.</source>
  <publisher>American Physical Society</publisher>
  <pubdate>2014</pubdate>
  <volume>112</volume>
  <fpage>200501</fpage>
</bibl>

<bibl id="B40">
  <title><p>Extracting quantum work statistics and fluctuation theorems by
  single-qubit interferometry</p></title>
  <aug>
    <au><snm>Dorner</snm><fnm>R</fnm></au>
    <au><snm>Clark</snm><fnm>S R</fnm></au>
    <au><snm>Heaney</snm><fnm>L</fnm></au>
    <au><snm>Fazio</snm><fnm>R</fnm></au>
    <au><snm>Goold</snm><fnm>J</fnm></au>
    <au><snm>Vedral</snm><fnm>V</fnm></au>
  </aug>
  <source>Phys. Rev. Lett.</source>
  <publisher>APS</publisher>
  <pubdate>2013</pubdate>
  <volume>110</volume>
  <issue>23</issue>
  <fpage>230601</fpage>
</bibl>

</refgrp>
} 

\end{backmatter}

\end{document}